\documentstyle[aps,prl,twocolumn,psfig]{revtex}
\draft
\newfont{\ssf}{cmss10 scaled 1000}

\def\ba{\begin{eqnarray}}

\def\bcal{\cal}
\def\CP{{\cal C_P}}
\def\dd{{\rm d}}

\def\ea{\end{eqnarray}}
\def\ee{\mbox{\rm e}\,}

\def\Ho{\mathop{\rm H{}}^{\,(1)}\nolimits}
\def\Ht{\mathop{\rm H{}}^{\,(2)}\nolimits}
\def\Hot{\mathop{\rm H{}}^{\,(1,2)}\nolimits}

\def\IM{\mathop{\Im{\rm m}}\nolimits}
\def\J{\mbox{\rm J}}

\def\lb{\left(}

\def\lg{\langle\:}

\def\ra{\rightarrow}

\def\rg{\:\rangle}

\def\rb{\right)}

\def\RE{\mathop{\Re{\rm e}}\nolimits}

\def\SI{S^{\,\text{(I)}}}
\def\SO{S^{\,\text{(O)}}}

\font\scal=cmsy9 at 12.truept
\begin{document}
\title{Dynamical Tunneling in Mixed Systems}
\author{S.\ D.\ Frischat and E.\ Doron$\,$\cite{byline}}
\address{Max--Planck--Institut f\"ur Kernphysik, Postfach 103980,
69029 Heidelberg, F.R.G.}               

\date{\today}

\twocolumn[
\maketitle
\widetext%
\vspace*{-6.45mm}
\leftskip=1.9cm\rightskip=1.9cm
\begin{abstract}
$\quad$
We study quantum-mechanical tunneling in mixed dynamical systems
between symmetry-related phase space tori separated by a chaotic
layer.  Considering e.g.~the annular billiard we decompose
tunneling-related energy splittings and shifts into sums over paths in
phase space.  We show that tunneling transport is dominated by {\em
chaos-assisted paths\/} that tunnel into and out of the chaotic layer
via the ``beach'' regions sandwiched between the regular islands and
the chaotic sea. Level splittings are shown to fluctuate on two scales
as functions of energy or an external parameter: they display a dense
sequence of peaks due to resonances with states supported by the
chaotic sea, overlaid on top of slow modulations arising from
resonances with states supported by the ``beaches''. We obtain
analytic expressions which enable us to assess the relative importance of
tunneling amplitudes into the chaotic sea vs.\ its internal transport
properties. Finally, we average over the statistics of the chaotic
region, and derive the asymptotic tail of the splitting distribution
function under rather general assumptions concerning the fluctuation
properties of chaotic states.
\vspace*{-1mm}
\vspace*{-1mm}
\end{abstract}
]
\narrowtext
\section{Introduction}

A detailed understanding of how the coexistence of classically regular
and chaotic phase space areas is reflected in the corresponding
quantum dynamics poses one of the challenging problems in the field of
``quantum chaos'' \cite{Bohigas90a}. Even though semiclassical
theories exist for the two limiting cases of fully integrable
\cite{EBK}, or fully chaotic classical
dynamics \cite{GutzwillerBook}, the quantum mechanical properties of
systems with ``mixed'' classical dynamics have up to date not been
amenable to a semiclassical formulation. The quest for such a theory
is importuned by the fact that mixed systems comprise the majority of
dynamical systems found in nature.

Out of the wealth of phenomena reported in mixed systems, a
particularly interesting one is genuinely quantum-mechanical in
nature: tunneling. A situation that has received much attention is
the one in which tunneling takes place between
distinct, but symmetry-related regular phase space regions
separated by a chaotic layer. Interest surged when it was discovered
that energy splittings can increase dramatically with
chaoticity of the intervening chaotic layer
\cite{LinBallentine,Bohigas93,Utermann94}. This 
was attributed to a suggested mechanism of {\em chaos-assisted
tunneling\/} \cite{Bohigas93,Bohigas93a,Tomsovic94} in which
tunneling takes place not  in a single tunneling transition, but in a
multi-step process containing tunneling transitions between regular
tori and the chaotic region, as 
well as chaotic diffusion inside the chaotic sea. Since a large part
of the phase space distance is thus traversed via classically allowed
transitions, indirect paths can be expected to carry considerably more
tunneling flux than direct ones.

Additional evidence was given by the observation that, apart from an
overall enhancement, the tunneling splittings vary rapidly
over many orders of magnitude as a function of energy, Planck's
constant $\hbar$ or other system\linebreak parameters. This was attributed to
the occurrence of avoided crossings between regular doublets and
chaotic states \cite{Bohigas93}, which made it possible to further
establish chaos-assisted tunneling by studying its effect on
statistical properties such as the splitting 
distribution function. Comparison with predictions of appropriate
random matrix models showed very good agreement
\cite{Bohigas90a,Tomsovic94,Leyvraz96}.
However, the lack of a 
semiclassical description of the tunneling processes remained as
a gap between the quantum and the classical picture, and --- more
importantly --- the
size of the tunneling amplitudes was unknown in the systems under
study, which made a {\em direct\/} and {\em quantitative\/} treatment
of the phenomenon impossible.  

Both of these problems were addressed in an earlier publication by the
authors of this work \cite{Doron95b} in which a semiclassical analysis
of tunneling 
processes in the annular billiard was performed, and a formula for the
contribution of chaos-assisted paths to the energy splitting was
derived. Here, we give a detailed account of our findings. Particular
emphasis will lie on the description of how the tunneling rate is
affected by  phase-space structures within the chaotic region, namely
the existence of an intermediate ``beach'' region sandwiched 
between classically regular islands and the chaotic sea.

The structure of this Paper is as follows. In Section
\ref{dyntunn:sect}, we review the basic ideas underlying this
work. Also, we introduce the model system under consideration, the
annular billiard. In Section \ref{quant:sect}, we introduce the method of
our analysis, the scattering approach to the quantization of closed
systems, and explicitly construct the ``scattering matrix'' $S$ for the
annular billiard. We then show in Section \ref{cat:sect} how the
scattering matrix approach can, under rather general assumptions, be
implemented to the study of tunneling in phase space. We explain how
$S$ can be approximated by a five-block matrix model with different
blocks representing regular dynamics on either of the islands, beach
motion close to each island, and chaotic dynamics in the center of the
chaotic sea. We derive formulas for $S$-matrix eigenphase shifts and
splittings in terms of paths passing through different combinations of
blocks, laying emphasis on the effects arising from the inclusion of
the beach blocks. Additionally, we track how tunneling flux spreads in
phase space and give a detailed discussion of the interplay of tunneling
probabilities into, and transport properties within the chaotic layer. 
Finally, in Section \ref{Statistics} we calculate 
statistical quantities --- such as the splitting distribution function
and median values for the splitting --- by averaging over the
properties of the chaotic block. We conclude with a discussion.

\section{Dynamical Tunneling}
\label{dyntunn:sect}
\subsection{Classical and Quantum Mechanics of Mixed Systems}

\subsubsection{Correspondence of Wavefunctions with Classical Structures}
\label{corresp:sect}
Systems with classically mixed dynamics display both regular and
chaotic behavior, depending on the starting conditions of the
trajectory considered. The structure of phase space can conveniently
be probed by use of a {\em Poincar\'e surface of section (PSOS)\/}
\cite{LichtenbergBook}, a phase space cut $\Gamma$  giving rise to the
{\em Poincar\'e map}
\begin{displaymath}
  (Q,P)_i\mapsto (Q,P)_f\ ,\quad 
  (Q,P)_{i/f}=(Q(\bbox{x_{i/f}}),P(\bbox{x_{i/f}}))\ ,  
\end{displaymath}
where $(Q,P)$ is a set of canonically conjugate variables, and
$\bbox{x}_{i/f}\in\Gamma$ are connected by the system dynamics. 
If one starts out with highly localized
distributions and plots iterates of the {\em Poincar\'e cell\/}
$\CP=\{(Q(\bbox{x}),P(\bbox{x})) : \bbox{x}\in\Gamma\}$, then chaotic
areas show up as areas which quickly become more or less uniformly
covered, while regular motion remains confined to lower-dimensional
manifolds on $\CP$. In a mixed
system, both types of structures appear, and one arrives at plots of
$\CP$ such as the one presented in
Fig.~\ref{smat:PSOS:fig} (see below).

In order to associate a system's quantum eigenstates $\psi$ with
classical features --- such as chaotic regions or regular tori --- one
often uses the {\em Wigner transformation\/} \cite{WignerHillery}
of the projector $|\psi\rangle\langle\psi|$. By smoothing over
minimal-uncertainty wavepackets one obtains the 
{\em Husimi distribution\/} \cite{HusimiTakahashi} that defines a
real, non-negative probability density in phase space. 
We will tacitly invoke the Wigner/Husimi concept in the
sequel when referring to the correspondence of quantum states with
phase space structures.

Until now, no general theory for the quantization of mixed systems is
at hand.  However, the understanding has emerged that, in the
semiclassical limit, quantum states can unambiguously be classified as
``regular'' and ``chaotic'' (for a review, see
\cite{Bohigas93}). Regular states are supported by classical tori
obeying EBK quantization rules \cite{EBK},
whereas chaotic states are associated with chaotic phase space regions
(or subsets of it \cite{Heller84}). The structure of chaotic states is
to date not fully understood and is presently the subject of intensive
research.  Classification of states as regular and chaotic can become
problematic at intermediate energy (or $\hbar$), since EBK-like
quantization rules can apply also to states residing on chaotic phase
space regions lying in close proximity to the regular island
\cite{Noid77,Bohigas90,Utermann94}.  Loosely speaking, the regularity
of classical islands can quantum-mechanically extend into the chaotic
sea, and states of an intermediate nature emerge.

\subsubsection{Effect of Phase Space Symmetries}

To discuss the effect of phase space symmetries on the structure of
quantum states, we consider a system with a discrete two-fold phase
space symmetry ${\cal T}$. We suppose that there are two disjoint
phase space objects ${\cal A}_1$ and ${\cal A}_2$, each of which is
invariant under the classical dynamics, mapped onto another by the
symmetry operation, ${\cal A}_2={\cal T}{\cal A}_1$. We also suppose that,
in the semiclassical limit, each of the ${\cal A}_{1/2}$ supports a
set of states primarily localized on it. Let us, for the sake of
definiteness, consider the case when the ${\cal A}_{1/2}$ are EBK-quantized
tori.  On each of the tori, one can construct {\em quasi-modes\/}
$\psi_r^{\,(1)}(\bbox{q})$ and
$\psi_r^{\,(2)}(\bbox{q})=\psi_r^{\,(1)}({\cal T}\bbox{q})$ that obey
the Schr\"odinger equation to any order of $\hbar$ \cite{MaslovBook}.
The corresponding EBK energy eigenvalues $E_r$ are then degenerate to
any order in $\hbar$. However, exact quantum states are constrained to
be symmetric or antisymmetric under ${\cal T}$,
\begin{eqnarray}
  \psi_r^{\,\pm}(\bbox{q})\approx\frac{1}{\sqrt 2}\left(
    \psi_r^{\,(1)}(\bbox{q})\pm\psi_r^{\,(2)}(\bbox{q})\right)\ ,
\label{tunn:evenodd}\end{eqnarray} and the energy degeneracy is lifted
by tunneling processes by an amount $\delta E_r$, giving rise to
tunneling oscillations with period $2\pi\hbar/\delta E_r$.

The best-known example of quantum-mechanical tunneling oscillations is
the one-dimensional symmetric double quantum well, where the phase
space symmetry ${\cal T}(x,p)=(-x,-p)$ connects regular tori in each
of the wells (for a careful discussion along the above line of
argument, see \cite{Tomsovic94}).  In systems of more than
two-dimensional phase space, symmetries can give rise to more 
intricate situations. The tori ${\cal A}_1$ and ${\cal A}_2$ must not
necessarily be separated by an energy barrier in configuration space,
but the transition from ${\cal A}_1$ to ${\cal A}_2$ can also be
forbidden by a {\em dynamical\/} law.  In this case, there is a
dynamical variable other than energy that is conserved by classical
dynamics, but violated by quantum dynamics
\cite{Miller7274,Miller75}.  The case of quantum doublets
connected by tunneling processes of this type was first reported by
Davis and Heller \cite{Davis81} who also coined the term {\em
dynamical tunneling\/}. A particularly clear example of dynamical
tunneling will be presented in Section \ref{annular:sect} in the
discussion of the annular billiard. As in the case of energy barrier
tunneling, splittings due to dynamical tunneling can be expected to be
very small, since classical transport from ${\cal A}_1$ to ${\cal
A}_2$ is forbidden.

Note that the formation of doublets is determined by the phase space
{\em topology\/} of the supporting region, not its regularity or
chaoticity.  The occurrence of doublets has also been observed in
situations, in which the localizing mechanism was due to dynamical
localization \cite{Bohigas90,Utermann94,Casati94} or scarring
\cite{Frischat97}. 
Conversely, a phase space structure ${\cal B}= {\cal T}{\cal B}$
mapped onto itself supports states that do not form 
doublets, regardless of its dynamical nature.

\subsection{Chaos-Assisted Tunneling}

Apart from the possibility of dynamical tunneling, tunneling processes
in systems of more than one degree of freedom can have an additional
aspect of interest:  the appearance of chaos in the region of phase
space traversed by the tunneling flux.  
As an early paradigm of such a system, Lin and Ballentine
\cite{LinBallentine} proposed the periodically
driven double well potential, where chaoticity can gradually be
introduced by increasing the driving strength. Lin and Ballentine
performed a numerical study of tunneling oscillations between states
associated with regular tori corresponding to classical motion
confined to either bottom of the well. They observed that, as the
separating phase space layer grows more chaotic with increasing driving
strength, tunneling rates are enhanced by orders of magnitude over
the rate in the undriven system (the integrable 1-$d$ double well). 
In a later study of the same system, Utermann {\em et al.\/}
\cite{Utermann94} established that the tunneling rate of a wave packet
initially localized on one regular island is determined
not by the wave packet's overlap with the other island,
but by its overlap with the chaotic sea, pointing at a
role for classically chaotic diffusion as a mediator of quantum
tunneling flux.

In a parallel and simultaneous line, Bohigas, Tomsovic and Ullmo
advocated the interpretation that the enhancement of tunneling was a
case of {\em resonant\/} tunneling due to the occurrence of avoided
crossings of the tunneling doublet's eigenenergies with the
eigenenergy of a state residing on the intervening phase space layer
\cite{Bohigas93}. For obvious reasons, the phenomenon was named {\em
chaos-assisted tunneling}.  The interpretation of tunneling
enhancement in terms of a three-level process was derived from the
observation that splittings of regular doublets are rapidly
fluctuating quantities as functions of parameters such as energy,
Planck's constant $\hbar$, or other model parameters --- much in
contrast to the smooth, at most oscillatory, dependence of tunneling
rates on $\hbar$, say, in the case of energy barrier tunneling
\cite{WilkinsonCreagh}.

Since a semiclassical description of tunneling matrix elements was
lacking, 
Bohigas and coworkers focussed on the statistical fingerprints
of chaos-assisted tunneling, with emphasis on the consequences of
resonance denominators on the splitting distribution. To this end, the
interaction of regular doublets with chaotic states was formulated in
terms of a block matrix model \cite{Bohigas93}, in which properties of
states residing on the chaotic sea were approximated by use of random
matrix theory \cite{BohigasSchool}.  This model was subsequently
refined by Tomsovic and Ullmo \cite{Tomsovic94} to take into account
the effect of additional time scales in the chaotic dynamics that can
appear when residual phase space structures, such as cantori, are
present in the chaotic sea acting as imperfect transport
barriers. Predictions made using these block-matrix models showed good
agreement with numerically calculated splitting distributions. Along
these lines, Ullmo and Leyvraz \cite{Leyvraz96} were also able to
derive analytic expressions for the splitting distributions in the
case of structure-less chaotic dynamics, as well as for a structured
chaotic sea. Again, theoretical predictions showed good agreement with
exact numerical data.

\subsection{The Annular Billiard}
\label{annular:sect}
We now introduce the specific system under consideration in this
work, the annular billiard. It was proposed by Bohigas {\em et
al.\/}~\cite{Bohigas93a} and consists of the area trapped between two
non-concentric circles of radii $R$ and $a<R$ centered at $(x,y)$
coordinates $O\equiv(0,0)$ and $O'\equiv(-\delta,0)$, respectively.
We consider the case of $\delta<a$ and set $R=1$, unless otherwise
stated. Note that the billiard is symmetric under reflections at 
the $x$-axis.

\begin{figure}
\protect\vspace*{.2cm}
\centerline{
\psfig{figure=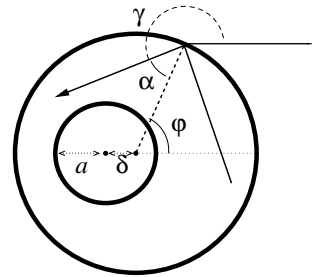,width=4cm,angle=-90}}
\vspace*{.5cm}
  \caption{Parameterization of classical trajectories. Note that 
    coordinates $(\gamma,L)$ are {\em not\/} the Birkhoff coordinates
    $(\varphi,L)$ usually employed in billiards. The two coordinate sets
    are related by $\varphi=\gamma+\alpha-\pi$, where
    $\alpha=\arcsin(L/R)$.}
  \label{bla}
\end{figure}

\subsubsection{Classical Dynamics}

Classical motion in a billiard is given as free flight between
specular reflections at the boundaries.  We select the PSOS $\Gamma$
as a circle of radius $r$ concentric with the outer circle and choose
$r$ to be infinitesimally smaller than one. Upon in-bound passage
through $\Gamma$ --- or, equivalently, after reflection from the outer
circle --- we record the trajectory's coordinates $(\gamma,L)$, where
$\gamma$ denotes the angle of the velocity vector with the $x$-axis
and $L=\sin\alpha$ is the classical impact parameter with respect to
$O$ (see Fig.~\ref{bla}).  $(\gamma,L)$ are canonically conjugate with
respect to the {\em reduced\/} action \cite{Miller75}, and the
Poincar\'e cell is given by $\CP=[0,2\pi]\times]-1,1[$. The billiard's
mirror symmetry $y\mapsto -y$ translates into an invariance of $\CP$
under the mapping $(\gamma,L)\mapsto(2\pi-\gamma,-L)$.

\begin{figure}
\protect\vspace*{.2cm}
\centerline{\psfig{figure=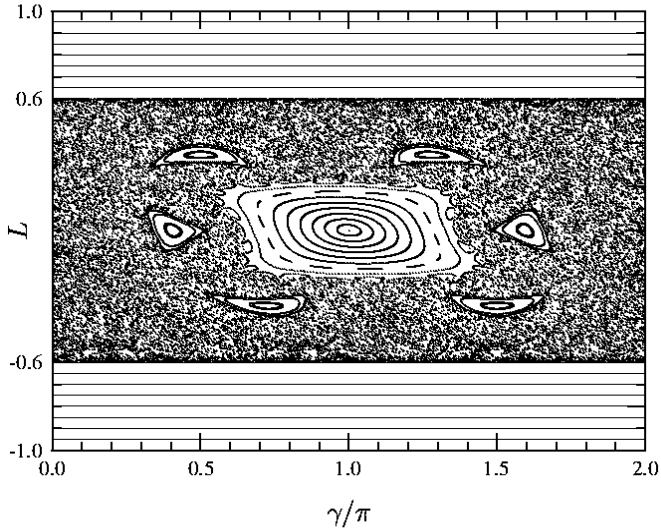,width=9cm}\vspace*{.2cm}}
  \caption{Poincar\'e plot of classical motion in the annular billiard at
    $a=0.4$ and $\delta=0.2$.}
  \label{smat:PSOS:fig} 
\end{figure}

In Fig.~\ref{smat:PSOS:fig} we present a Poincar\'e plot of $\CP$ at
parameter values $a=0.4$ and $\delta =0.2$. Rays of $|L|>a+\delta$ do
not hit the inner circle, but forever encircle the inner disc at
constant $L$ filling horizontal strips (or subsets of horizontal
strips) in the Poincar\'e plot.  Each of these {\em whispering gallery
(WG)\/} tori, specified by its impact parameter $L$, is associated with
a partner torus $-L$ by the mirror symmetry.  These WG tori will be the
tunneling tori under consideration in this work.

Rays of intermediate impact parameter $|L|<a+\delta$ will eventually
hit the inner circle, and since angular momentum is then not
preserved, motion is no longer integrable.  This can give rise to the
whole range of phenomena associated with non-integrable systems of
mixed phase space: regular islands and island chains, chaotic regions,
partial transport barriers (cantori) and the like.  The structure of
the phase space layer $|L|<a+\delta$ is primarily organized by two
fixed points of the Poincar\'e map: (i) an unstable fixed point at
$(\gamma,L)=(0,0)$ and its stable and unstable manifolds, along which
a chaotic region spreads out, and (ii) a stable orbit
$(\gamma,L)=(\pi,0)$ at the center of a regular island of
``libration'' trajectories. The fixed points correspond to rays along
the symmetry axis on the left hand side and on the right hand side of
the inner circle, respectively.

It will turn out to be of great importance that there is a  region of
chaotic, but relatively stable motion surrounding each regular island.
In the strip of $|L|\lesssim a+\delta$ this stability is easy to
understand, as trajectories typically encircle the inner disc many
times until a hit occurs, and at each hit the change in impact
parameter is small. The ``stickiness'' of this beach region is increased
by the presence of regular island chains and of cantori that are the
remains of broken WG tori.

\subsubsection{Quantum Mechanics}

Quantum mechanics of the annular billiard with Dirichlet boundary
conditions is given by the Helmholtz equation 
\[
  (\Delta+k^2)\,\psi(\bbox{q})=0
\]
and the requirement of vanishing wavefunction on the two circles.
The wave number $k$ is related to energy by $E=\hbar^2k^2/2m$. 
(We note that there exists an analogy between quantum billiards and
quasi-two-dimensional microwave resonator which has proven
instrumental in many experimental realizations of billiard systems
\cite{Richter96}.) 

We give here only a qualitative picture of the quantum states,
deferring a full solution to Section 
\ref{annular:2}. It is most appropriate to decompose the wavefunction into
angular momentum components by writing
\begin{eqnarray} 
  \psi(r,\varphi) \!= \!\sum_{n=-\infty}^{\infty}\!i\,^n\left[ \alpha_n
    \mathop{\rm H{}}^{\text{\,(2)}}\nolimits_n(kr) + \beta_n \mathop{\rm
      H{}}^{\,(1)}\nolimits_n(kr)\right] \text{e}^{in\varphi} \ ,
  \label{smat:wavefn}
\end{eqnarray}
where $(r,\varphi)$ are polar coordinates with respect to $O$, and
$\mathop{\rm H{}}^{\,(1,2)}\nolimits_n(x)$ denote Hankel functions of
the first and second kind, respectively, of order $n$. We recall that
angular momentum quantum numbers $n$ are in the semiclassical limit
related to classical impact parameters $L=n/k$. 
 
To understand the nature of quantum states supported by the annular
billiard, it is instructive to first consider the concentric billiard
and then to ``turn on'' the eccentricity $\delta$. If 
$\delta=0$, then angular momentum is conserved, and states are
paired in energetically degenerate doublets composed of angular
momentum components $n$ and $-n$.  In the eccentric system
($\delta\neq 0$), the degeneracy is lifted by the breaking of
rotational invariance. However, angular momentum doublets are affected
to different degrees --- depending on the size of $n$ relative to
$k(a+\delta)$. The symmetry breaking has large effect on doublets of
small angular momentum $|n|<k(a+\delta)$ corresponding to classical
motion that can hit the inner circle. For low-$n$ doublets, the
doublet pairing disappears quickly with increasing $\delta$ and
``chaotic'' eigenstates appear that spread out in angular momentum
components roughly between $-k(a+\delta)$ and $k(a+\delta)$. High-angular
momentum doublets with $|n|>k(a+\delta)$ are affected only
little by the symmetry breaking. The doublet-pairing persists, and
energy degeneracy is only slightly lifted. States are primarily
composed of symmetric and antisymmetric combinations of $n$ and $-n$
angular momentum components,
\begin{equation}
   \mid \bbox{\alpha}^{\,(\pm)} \rg \approx
\frac{1}{\sqrt{2}}\big(\mid n\rg \pm \mid -n \rg\big)\ .
\end{equation}
Note that each of the quasi-modes $|\pm n\rangle$ corresponds to
classical motion on the WG torus $\pm L$.

\begin{figure}
\psfig{figure=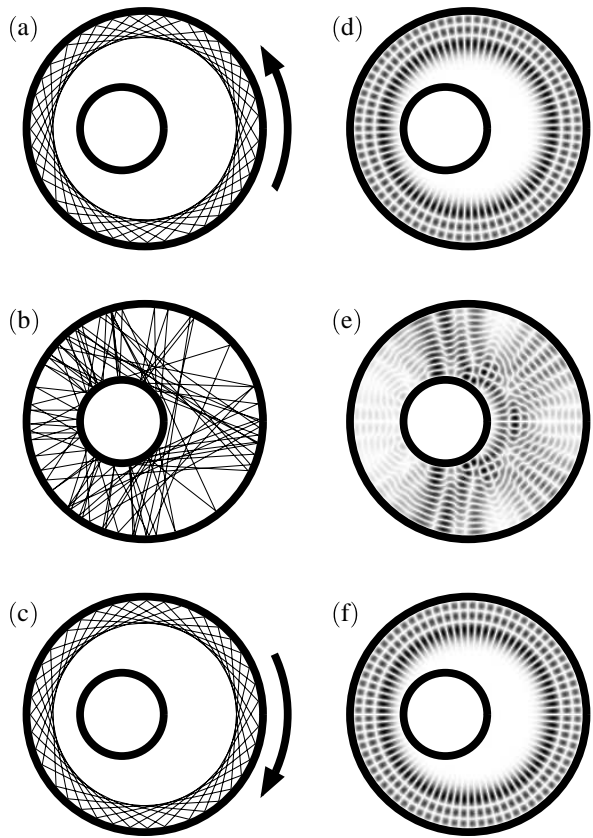,width=9cm}
  \caption{Comparison of classical dynamics and quantum states of the 
    annular billiard.
    (a--c) Classical motion: (a,c) regular trajectories of $\pm L$,
    $|L|>a+\delta$,  
    (b) chaotic trajectory of $|L|<a+\delta$.
    (a--c) Eigenmodes: (a,c) doublet of ``regular'' eigenstates at
    $k\approx54.434$, and (b) ``chaotic'' eigenstate at $k\approx 60.252$.}
  \label{compar:fig}
\end{figure}

We present plots of a ``regular'' doublet quantized at $k\approx 55$
and a ``chaotic'' state at $k\approx60$ in Fig.~\ref{compar:fig} and
compare them to classical trajectories with starting conditions in
the chaotic sea and on WG tori, respectively. The
correspondence between quantum states and the nature of classical
dynamics is clearly visible.
 
\subsubsection{Tunneling Between Whispering Gallery Tori}

Let us discuss the high-angular momentum doublets in more detail. As
explained above, the energy splitting $\delta
E_n$ between $| \bbox{\alpha}^{\,(+)} \rangle$ and $|
\bbox{\alpha}^{\,(-)} \rangle$ gives rise to tunneling oscillations
between quasi-modes $|\pm n \rangle$ associated with WG tori $\pm L=\pm
n/k$ $(L>a+\delta)$. A quantum particle prepared in state $|n\rangle$
will therefore change its sense of rotation from counter-clockwise to
clockwise and back to counter-clockwise with period $2\pi\hbar/\delta
E_n$. Note that this tunneling process serves as a particularly clear
example of dynamical tunneling. It occurs in {\em phase space\/}
rather than configuration space, as the corresponding tori are
identical in configuration space. Also, the tunneling process does not
pass under a potential barrier in configuration
space. In fact, energy does not play any role in the tunneling, as
energy is related only to the absolute value of the momentum vector
and not to its direction. Rather,
the tunneling process violates the {\em dynamical\/} law of classical
angular momentum conservation for rays of large impact parameter.

The concept of chaos-assisted tunneling can be nicely visualized for
the case of tunneling between WG modes in the annular billiard. In
fact, the annular billiard was proposed as a paradigm for
chaos-assisted tunneling in Ref.~\cite{Bohigas93a}.  In chaos-assisted
processes, tunneling tori $\pm L$ are connected not by direct
transitions between $n\mapsto -n$, but by multi-step transitions
$n\mapsto\ell\ldots -\ell'\mapsto -n$.  A particle tunnels from $n$ to
some $\ell\lesssim k(a+\delta)$, traverses the chaotic phase space
layer by classically allowed transitions to reach the opposite side of
the chaotic sea $-\ell'\gtrsim -k(a+\delta)$, and finally tunnels
from there to $-n$. To establish this notion, Bohigas {\em et al.\/}
\cite{Bohigas93a} 
checked the behavior of splittings as the eccentricity is changed to
make the intervening phase space layer more chaotic. In a numerical
study, they compared the splittings of regular doublets with the rate
of classical transport across the chaotic layer.  The findings showed
that the splittings increase dramatically over many orders of
magnitude as chaotic transport becomes quicker.

However, without a quantitative --- possibly semiclassical --- theory,
it is impossible to separately analyze the importance of tunneling
amplitudes and transport properties of the intervening chaotic
layer. Usually, the parameter governing symmetry breaking in a mixed
system changes {\em both\/} tunneling amplitudes from/into the regular
torus and chaoticity in the intermediate layer.  In order to separate
the relative importance of these effects, a quantitative understanding
of the tunneling processes must be obtained. Such a quantitative
description of chaos-assisted tunneling was given in
Ref.~\cite{Doron95b} and will be developed in full detail in the sequel.

\section{Quantization by Scattering}
\label{quant:sect}
\subsection{General Description of the Method}

In this work, we will employ an scattering approach to quantization 
\cite{Doron92b,Bogomolny92b} which, in essence,
is constructed as the quantum-mechanical analogue of the classical
Poincar\'e surface of section method. For the sake of
self-containedness, we give a brief review of the
scattering method. 

Let us consider the case of a 
billiard ${\cal G}$ and introduce a Poincar\'e cut $\Gamma$ in
configuration space, thereby dividing ${\cal G}$ into two parts,
${\cal G}_+$ and ${\cal G}_-$. We suppose that $\Gamma$ can be chosen
along a coordinate axis (the $q_2$-axis, say) and that the wave
problem is separable on an infinitesimal strip around $\Gamma$. At a
given energy, one chooses a complete set of functions
$\phi_n^{(2)}(q_2)$ along $\Gamma$ and writes the wavefunction on an
infinitesimal strip around $\Gamma$ as
\begin{eqnarray}
  \psi(\bbox{q})=\sum_n\lb \alpha_n\phi_n^{(1,-)}(q_1) +
  \beta_n\phi_n^{(1,+)}(q_1)\rb \phi^{(2)}_n(q_2) \ .
  \label{scatt:wavefn}\end{eqnarray}
$\phi_n^{(1,+)}$ and $\phi_n^{(1,-)}$ are functions that, in the
semiclassical limit, correspond to waves traversing $\Gamma$ in
positive and negative $q_1$-direction, respectively. 
We assume that the set $\phi_n^{(2)}(q_2)$ is chosen such that quantum
numbers $n$ correspond to values $k^{(2)}_n$ of longitudinal wave
number. Then, $k_n^{(2)}$ and transverse wave numbers $k_n^{(1)}$ are
related by  $E_n=\hbar^2[(k_n^{(1)})^2+(k_n^{(2)})^2]/2m$.
Note that orthogonality of the modes on $\Gamma$ is ensured by the
choice of the $\phi_n^{(2)}$. Each of the
domains ${\cal G}_\pm$ constitutes a scattering system that
scatters waves $\phi_n^{(1,+)}\,\phi_n^{(2)}$ into waves
$\phi_n^{(1,-)}\,\phi_n^{(2)}$ and {\em vice versa\/}.  Associated
with these scattering systems ${\cal G}_\pm$ are scattering matrices
$S_\pm(E)$ that relate the coefficient vectors
\begin{eqnarray}
  \bbox{\beta} \:=\:S_-(E)\,\bbox{\alpha}\qquad\text{ and }\qquad
  \bbox{\alpha}\:=\:S_+(E)\,\bbox{\beta}\ .
  \label{scatt:scatt}\end{eqnarray}
The quantization condition is equivalent to the requirement of
single-valuedness of the wave function on $\Gamma$, and so to
the equivalence of the two scattering conditions. The system 
supports an eigenstate whenever the product matrix 
\[
  S(E)\equiv S_-(E)\,S_+(E) 
\]
has an eigenvalue of unity, hence the quantization condition reads
\begin{eqnarray}
  \det(S(E)-1)=0\ .
\label{poin:secular}\end{eqnarray}
At a quantized energy,
the wave function can be reconstructed from the corresponding
eigenvector $\bbox{\alpha}$ of $S$ via
Eqs.~(\ref{scatt:wavefn},\ref{scatt:scatt}). 
In principle, $S$ is an infinite-dimensional 
matrix. However, in many cases of interest one can choose the
$\phi_n^{(1,+)}\,\phi_n^{(2)}$ so that in the region of classically
allowed motion, the contribution $|\phi_n^{(1,\pm)}(q_1)|$ is
exponentially small for all but a finite number of indices (the
so-called ``open channels''). This allows the truncation of $S$ to
finite dimension, say $|n|\le\Lambda$, with an error that is
exponentially small. Both
scattering matrices $S_\pm$ can be constructed in a representation
such that they are unitary in the space of open modes
and, if the system is time-reversal invariant, symmetric. 
$S$ on the other hand is unitary, but not necessarily
symmetric.  In spite of this, we will in this paper also refer to $S$
as a scattering matrix.

It is clear by construction that $S$ is the quantum-mechanical
analogue of the Poincar\'e mapping \cite{Blumel90}. Its $N$-th iterate
$S^N$ constitutes a time-domain-like propagator. Note that the
iteration count $N$ of the Poincar\'e map does not correspond to a
stroboscopic discretization of time, but rather to a {\em
  fictitious\/} discrete time, since generally the time elapsed
between passages of $\Gamma$ can vary. If, however, we select our
surface of section properly, then return times will not vary by too
much, and $S$ will not be too different from the genuine time-domain
propagator.

\subsection{Scattering Matrix of the Annular Billiard}
\label{annular:2}

It is fairly straightforward to implement the scattering approach to
the case of
the annular billiard. As discussed in Section \ref{annular:sect}, we
choose $\Gamma$ as a circle of radius $r$, where $a+\delta<r\lesssim
1$. Since
classical impact parameter is conserved by motion on the WG
tori, we choose $q_2=\varphi$ and $\phi_n^{\,(2)}=\exp(in\varphi)$ on
$\Gamma$. Waves traversing $\Gamma$ are given by outgoing and ingoing
cylinder waves $\phi_n^{\,(1,+)}=i^n\Ho_n(kr)$,
$\phi_n^{\,(1,-)}=i^n\Ht_n(kr)$, and we obtain the decomposition in
Eq.~(\ref{smat:wavefn}), 
\begin{eqnarray} 
  \psi(r,\varphi) \!=\! \sum_{n=-\infty}^{\infty}i^n\left[ \alpha_n
    \mathop{\rm H{}}^{\text{\,(2)}}\nolimits_n(kr) + \beta_n \mathop{\rm
      H{}}^{\,(1)}\nolimits_n(kr)\right] \text{e}^{in\varphi} \ .
\label{smat:wavefn2}\end{eqnarray}

In the present
example, outgoing waves are scattered to ingoing waves by the interior
of the outer circle --- giving rise to the scattering condition
\mbox{$\bbox{\alpha}=S^{\,\text{(O)}}(k)\,\bbox{\beta}\:$} --- and 
ingoing waves are reflected off
the exterior of the inner circle, which leads to the
relation \mbox{$\bbox{\beta}=S^{\,\text{(I)}}(k)\,\bbox{\alpha}\:$}. 
In order to derive the explicit formulas for $\SO$ and $\SI$, we note
that the Dirichlet boundary condition
$\psi(R,\varphi)=0$ on the outer circle leads to
\begin{equation}
  \SO_{n,m}(k)=-\frac{\Ho_n(kR)}{\Ht_n(kR)}\:\delta_{n,m}\ .
  \label{smat:external}
\end{equation}
$\SO(k)$ is diagonal, in accordance with angular momentum
conservation in scattering events off the outer circle. 

$\SI(k)$ is derived by performing a coordinate change to the 
primed coordinates defined with respect to $O'=(\delta,0)$. In this
representation the scattering matrix is 
$S^{\,\prime\,\text{(I)}}(k)=-\Ht_n(ka)/\Ho_n(ka)\,\delta_{n,m}$. 
Transformation back to unprimed coordinates is done by the addition
theorem for Bessel functions (see e.g.~\cite{Abramowitz})
\[
  \Hot_n(kr)\,\ee^{in\varphi}=\sum_{\ell=-\infty}^\infty 
  \J_{n-\ell}(k\delta)\, \Hot_\ell(kr')\,\ee^{i\ell\varphi'}
\]
for $\delta<a=r'$. $\J_n(x)=(\Ho_n(x)+\Ht_n(x))/2$ denotes the
Bessel function of order $n$. We arrive at
\begin{displaymath}
  \SI_{n,m}(k)=-i^{(n-m)} \sum_{\ell=-\infty}^\infty  \mathop{\rm
    J{}}\nolimits_{n-\ell}(k\delta)\, 
    \mathop{\rm J{}}\nolimits_{m-\ell}(k\delta)\:
    \frac{\mathop{\rm H{}}^{\text{\,(2)}}\nolimits_\ell(ka)}
    {\mathop{\rm H{}}^{\,(1)}\nolimits_\ell(ka)} \;. 
\end{displaymath}
Time-reversal invariance imposes the symmetry\linebreak
$\SI_{m,n}(k)=(-)^{(n-m)}\,\SI_{n,m}(k)$ on $\SI(k)$ and by virtue of the
mirror symmetry $\SI_{-n,-m}(k)=\SI_{n,m}(k)$. (Note that the formula
given for $\SI(k)$ differs by the factor $i\,^{(n-m)}$ from a
formula given earlier by us \cite{Doron95b} and consequently, also
the symmetries are different. This is due to a slightly different
choice of basis in (\ref{smat:wavefn2}).) The full $S$-matrix is
then given as
\begin{eqnarray}
  &&S_{n,m}(k)=\nonumber\\
  &&i\,^{(n-m)}\frac{\Ho_n(kR)}{\Ht_n(kR)}\,
  \sum_{\ell=-\infty}^\infty\mathop{\rm
    J{}}\nolimits_{n-\ell}(k\delta)\, 
  \mathop{\rm J{}}\nolimits_{m-\ell}(k\delta)\:
  \frac{\mathop{\rm H{}}^{\text{\,(2)}}\nolimits_\ell(ka)}
  {\mathop{\rm H{}}^{\,(1)}\nolimits_\ell(ka)} \;.  
  \nonumber\end{eqnarray}

\begin{figure}
\centerline{\protect\vspace*{-4cm}\protect\hspace*{.5cm}
  \psfig{figure=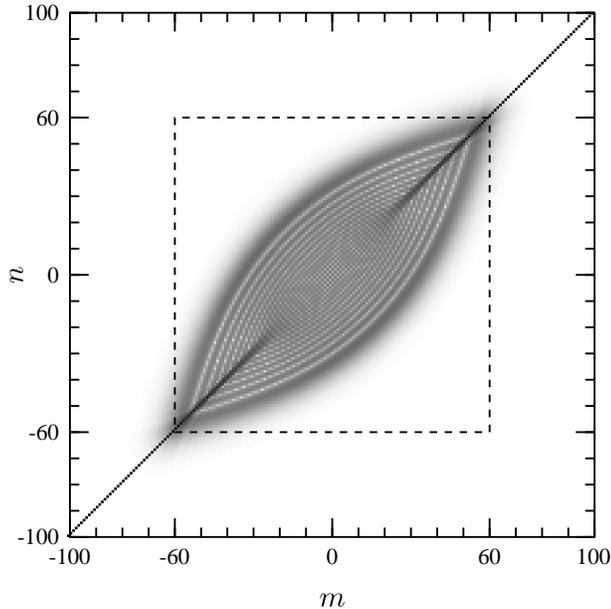,width=10cm}
  \protect\vspace*{2.2cm}\protect\hspace*{2cm}}
  \caption{Grayscale plot of the quantum $S$-matrix $|S{_n,m}|$
    (arbitrary gray scale). The dashed line indicates the region of
    classical angular momentum mixing $|n|<k(a+\delta)$.}
  \label{Smat:fig}
\end{figure}
Using the relation $\Hot_{-n}(x)=(-)^n\Hot_n(x)$ for integer $n$, one 
verifies that the spatial symmetry of the annular billiard translates
into the $S$-matrix symmetry
\begin{eqnarray}
  S_{-n,-m}\:=\:S_{n,m}\ .
  \label{smat:smatsymm}\end{eqnarray}
For eigenvectors $\bbox{\alpha}^{(j)}$ of $S$,
\begin{eqnarray}
  \alpha_{-n}^{(j)} ={\sigma_j}\, \alpha_n^{(j)}\ ,
  \label{smat:vecsymm}\end{eqnarray}
where $\sigma_j\in\{\pm 1\}$. Whenever the system supports an
eigenstate, $\sigma_j$ determines the symmetry of the corresponding
wave function with respect to the $x$-axis.

In Fig.~\ref{Smat:fig}, we show a grayscale plot of $|S_{n,m}|$ as a
function of ingoing and outgoing angular momentum for the parameter
values $a=0.4$, $\delta=0.2$, $k=100$. The overall structure
of $S$ is governed by classically allowed transitions: it is
mainly diagonal in the region of high angular momentum
$|n|,|m|>k(a+\delta)$, whereas the inner block reflects the dynamics given by
the classical deflection function. Here, the main amplitude is
delimited by two ridges that correspond to classical rainbow
scattering.

\begin{figure}
\protect\vspace*{-.9cm}
\centerline{\protect\hspace*{0cm}
  \psfig{figure=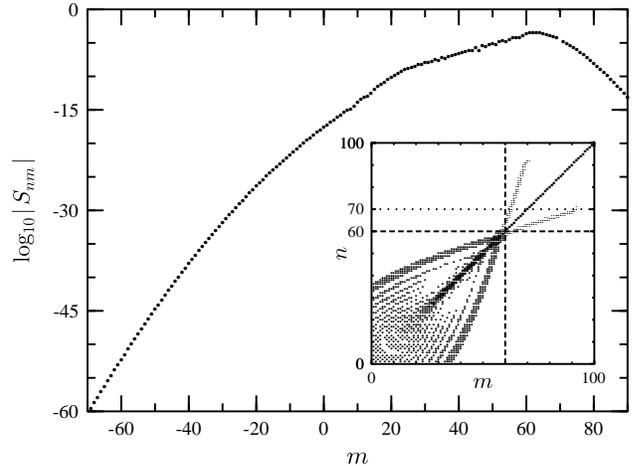,width=12cm}}
  \protect\vspace*{-.8cm}
  \protect\caption{Absolute values of tunneling matrix elements
    $|S_{n,m}|$ for $n=70$ and $m=-100,\ldots,100$ (logarithmic scale).
    Inset: first quadrant of $S$, showing the tunneling ridges. 
    (Taken from Ref.~\protect\cite{Doron95b}.)}
  \label{tunnel:fig}
\end{figure}

The tunneling amplitudes relevant to the WG splitting are contained in
$S$ as non-diagonal entries $S_{n,m}$ with $n>k(a+\delta)$.
Fig.~\ref{tunnel:fig} depicts $|S_{n,m}|$
for the above parameter values and $n=70$ (at $k(a+\delta)=60$). 
The tunneling amplitudes are largest ($\sim 10^{-4}$) around $m=63$ and
fall off faster than exponentially away from this maximum. As can be
seen in the inset of Fig.~\ref{tunnel:fig}, the line of
maximal tunneling amplitudes continues the line of rainbow ridges
into the regime of classically forbidden transitions. Close to these
tunneling ridges, and in the direction away from the diagonal, one can
observe oscillations in $|S_{n,m}|$ that resemble the Airy oscillations
well-known in diffraction theory \cite{Berry72}.  
It is interesting to note that the behavior of the
tunneling probabilities is {\em not\/} a monotonous function of
$|n-m|$, that is, of the phase space distance traversed. A mere
{\em distance} in phase space can therefore not serve to estimate the
behavior of tunneling probabilities.

We finally mention that the tunneling transitions considered here 
can be given a
semi-classical interpretation in terms of complex rays that interact
with the analytical continuation of the inner circle to complex
configuration space \cite{Doron95b,SteffenThesis,us_later}. 
These rays can either
scatter off the inner complex circle by a generalization of specular
reflection, or they can creep along a complexified inner circle (of
complex radius determined by the poles of the internal scattering
matrix)  by a mechanism similar to that
proposed by Franz \cite{Franz54} and Keller \cite{Keller62}.  
Every tunneling pair $n_i,n_f$ is connected by at least one complex
reflection trajectory with real initial (final) impact parameter
$L_i=n_i/k$ $(L_f=n_f/k)$ and complex initial (final) angle $\gamma_i$
$(\gamma_f)$.  Complex reflection and creeping trajectories give rise
to contributions to the tunneling matrix elements that scale with $k$
like
\begin{eqnarray}
  S_{\text{reflection}} 
  &\sim& k^{-1}\:\exp\Big(-k A_{\text{ref}} (i,f)\:\Big)\ ,
  \nonumber\\ 
  S_{\text{creeping}} 
  &\sim& k^{-2/3}\:\exp\Big(-k A_{\text{cr}} (i,f)
  + k^{1/3} B_{\text{cr}}(i,f)\:\Big)\ ,
  \nonumber\end{eqnarray}
respectively, where $A_{\text{ref}}$, $A_{\text{cr}}$ and
$B_{\text{cr}}$ are classical properties of the associated
trajectories that are independent of $k$. The relative importance of
the two contributions depends on $k$, the billiard geometry, and the
initial and final angular momenta considered. At the present parameter
values, the contribution due to reflected rays usually dominates the
one arising from creeping rays.

The semiclassical picture provides an intuitive explanation to the
tunneling ridges mentioned above: they can be identified as
combinations $n,m$ where one of the angles
$\gamma_i,\gamma_f$ is closest to reality. Also, the nearby oscillations
can be understood as arising from a coalescence of
two reflection saddle points.

\section{Treatment of Chaos-Assisted Tunneling}
\label{cat:sect}
\subsection{Implementation of the Scattering Approach}

We now discuss how the scattering approach can be implemented to
the treatment of chaos-assisted tunneling.  Let us suppose that the
system under consideration has a phase space symmetry of type
$(q_2,p_2)\mapsto(-q_2,-p_2)$, and that the $q_2$-axis can be chosen
as a PSOS $\Gamma$. We assume that the wave problem is locally
separable around $\Gamma$, which renders the problem tractable by the
scattering approach. 
Let us now pick two classical objects ${\cal A}_{1}$ and ${\cal
  A}_2={\cal T}{\cal A}_1$ and consider quasi-modes
$\psi_r^{\,(1/2)}(\bbox{q})$ supported by ${\cal A}_{1/2}$, as
discussed in Section \ref{dyntunn:sect}. We assume that motion on
${\cal A}_1$ and ${\cal A}_2$ corresponds to conservation of $p_2$ and
$-p_2$, respectively, and that $\pm p_2$ is semiclassically related to
the quantum number $\pm n$.
(Note that in a general system, the choice of the PSOS $\Gamma$ and a
proper basis for the scattering matrix can be a very difficult task. 
In this Paper, we will not deal with the problem of solving a general
scattering problem --- except for the annular billiard --- and assume
the scattering matrix $S_{n,m}$ to be known.)
By symmetry $p_2\mapsto-p_2$, we find that $S_{n,n}=S_{-n,-n}$, and
by virtue of the classical conservation of $p_2$, $S_{n,n}$ is
almost unimodular. The deviation of $|S_{n,n}|$ from unity will 
be due to classically forbidden (i.e.\ tunneling) transitions, and so
is expected to be small.

Let the quantization energies of the doublet be denoted by $E_n^\pm$,
and let $|\bbox{\alpha}^\pm(E_n^\pm)\rangle$ be the 
eigenvectors corresponding to the two quantum states. We now make the
approximation that the properties of these two vectors
are, to good precision, given by the eigenvector doublet
$|\bbox{\alpha}^\pm(E)\rangle$ at {\em one\/} fixed energy $E$ lying
between $E_n^+$ and $E_n^-$. The corresponding (generally non-zero)
eigenphases be denoted by $\theta_n^\pm(E)$.

Dropping the energy variable $E$, we decompose the
eigenphases $\theta_n^\pm$ in the form
\begin{eqnarray}
  \theta_n^\pm\:\equiv\:\theta_n^{\,(0)} + \Delta\theta_n^{\,(R)}
  +i\Delta\theta_n^{\,(I)}\pm \frac{1}{2}\,\delta\theta_n \ ,
  \label{shift:phase}
\end{eqnarray}
where $\Delta\theta_n^{\,(R,I)}$ are taken to be real, and
$\theta_n^{\,(0)} =-i\log S_{n,n}$.  The quantities
$\Delta\theta_n=\Delta\theta_n^{\,(R)} +i\Delta\theta_n^{\,(I)}$ and
$\delta\theta_n$ can be interpreted as the shift and the splitting,
respectively, of the exact eigenphases due to tunneling processes.
These eigenphase quantities are trivially related to the energy shift
and splitting, as will be explained below, and it is therefore
sufficient to calculate
$\Delta\theta_n$ and  $\delta\theta_n$. Note that the $\theta_n^\pm$
are real by unitarity of $S$, and therefore also
$\theta_n^{\,(0)}+\Delta\theta_n$ and $\delta\theta_n$ are real
quantities. We also note that 
\begin{eqnarray}
  \left[S^N\right]_{\pm,\pm}&\equiv&
    \langle\bbox{\alpha}^\pm |S^N|\bbox{\alpha}^\pm\rangle \nonumber\\
    &=&\ee^{iN\theta_n^{\,(0)}}\, \exp\lb iN\left[\Delta\theta_n 
    \pm \frac{1}{2}\,\delta\theta_n\right]\rb
  \label{shift:iter:phase}
\end{eqnarray}
for any integer $N$.

{}From the eigenvector doublet $|\bbox{\alpha}^\pm\rangle$ we can
now obtain the vector equivalent of quasi-modes  
\begin{eqnarray}
  |\pm\bbox{n}\rangle\:\equiv\:\frac{1}{\sqrt{2}} 
  \Big(\,\mid\bbox{\alpha}^+\:\rangle \pm 
         \mid\bbox{\alpha}^-\:\rangle\,\Big)\ .
\label{shift:quasi}\end{eqnarray}
It is clear that $|\bbox{n}\rangle$ and $|-\bbox{n}\rangle$ are 
localized at (or around) components $\pm n$, respectively.
Using the symmetry of $S$ and $|\bbox{\alpha}^\pm\rangle$, we write
\begin{equation}
  \left[S^N\right]_{+,+}\pm\left[S^N\right]_{-,-} 
    \:=\: 2\,\langle\:\bbox{n}\mid S^N\mid\pm\bbox{n}\:\rangle\ .
  \label{shift:iterates}
\end{equation}

In order to derive formulas for $\Delta\theta_n^{\,(R)}$ and
$\delta\theta_n$ that relate these quantities to matrix elements of
$S^N$, we now combine Eqs.~(\ref{shift:iter:phase}) and
(\ref{shift:iterates}).  
We choose a positive $N$ which satisfies
\begin{eqnarray}
  N \ll|\Delta\theta_n\pm\delta\theta_n/2|^{-1}
  \label{shift:split:con}\end{eqnarray}
and expand the second
exponential in Eq.~(\ref{shift:iter:phase}) to first order.
Considering upper signs in Eq.~(\ref{shift:iterates}) and taking
imaginary parts we obtain 
\begin{eqnarray}
  \Delta\theta_n^{\,(R)} \approx \frac{1}{N}\,\IM
  \Big\{
  \ee^{-iN\theta_n^{\,(0)}}\:
  \langle\:\bbox{n}\mid S^N\mid\bbox{n}\:\rangle
  \Big\}\ .
\label{shift:shiftSN1}\end{eqnarray}
Similarly, taking lower signs in  Eq.~(\ref{shift:iterates}) gives
\begin{eqnarray}
  \delta\theta_n \approx \frac{2}{N}\,\IM
  \Big\{
     \ee^{-iN\theta_n^{\,(0)}} \:
     \langle\:\bbox{n}\mid S^N\mid-\bbox{n}\:\rangle
     \Big\}\ .
\label{shift:splittingSN1}\end{eqnarray}

It is instructive to rephrase Eq.~(\ref{shift:splittingSN1}) for the
splitting:  starting from 
Eqs.~(\ref{shift:iter:phase},\,\ref{shift:iterates}) one can also 
contract the exponentials to a sine and arrive at
\begin{eqnarray} 
  \Big|\:\sin\lb\frac{N\delta\theta_n}{2}\rb\:\Big|
  \:=\: \left|\langle\:\bbox{n}\mid S^N\mid-\bbox{n}\rangle\right|\ ,
\label{shift:alternative}\end{eqnarray}
which has the form of tunneling
oscillations in time, $|\sin(\delta Et/2\hbar)|=
|\langle \bbox{n}|\exp(-iHt/\hbar)|-\bbox{n}\rangle|$, 
 with iteration count $N$ taking the role of time
and eigenphase splitting taking the role of energy splitting.  By
considering Eq.~(\ref{shift:splittingSN1}) we therefore probe the onset of
tunneling oscillations in the linear regime.

Use of Eqs.~(\ref{shift:shiftSN1},\ref{shift:splittingSN1}) for low $N$
will however necessitate precise knowledge of the eigenmodes
$|\pm\bbox{n}\rangle$ (see e.g.~\cite{Hackenbroich97b} for a recent
application of a similar formula for $N=1$). However, to exponential
precision, eigenvectors may be just as difficult to obtain
as the shift or the splitting itself. It is therefore important to
realize that use of Eqs.~(\ref{shift:shiftSN1},\ref{shift:splittingSN1})
for {\em large\/} $N$ may allow to extract the quantities of interest
using much less precise eigenvector information. Instead, one then
uses {\em dynamical\/} information --- in the framework of
time-domain like propagation with $S$ --- which will eventually allow
the interpretation of tunneling processes in terms of sums over paths
in phase space. To this end, let us reformulate
Eq.~(\ref{shift:quasi}) by writing  
\begin{eqnarray}
  \mid\bbox{\alpha}^\pm\,\rangle =\frac{1}{\sqrt{2}}\Big(\,\mid
  n\:\rangle \pm \mid -n \:\rangle\,\Big) + \sum_m\kappa^\pm_m \mid m
  \:\rangle\ ,
\label{shift:vector}\end{eqnarray}
where the $\kappa^\pm_m$ are expected to be small and $\kappa^\pm_{-m}
=\pm\kappa^\pm_m$, according to the symmetry of
$|\bbox{\alpha}^\pm\rangle$. The right-hand side of
Eq.~(\ref{shift:iterates}) then reads
\begin{equation}
   2 \langle\:\bbox{n}\mid S^N \mid \pm\bbox{n}\:\rangle
    \:=\: 2\,\left[S^N\right]_{n,\pm n} + {\cal C}_n^{\,(N,\pm)}\ .
  \label{shift:iterates2}
\end{equation}
Here, $[S^N]_{n,\pm n}$ denotes a matrix element of the $N$-th iterate
of $S$, and
${\cal C}_n^{\,(N,\pm)}={\scal c}_n^{\,(N,+)}\pm {\scal c}_n^{\,(N,-)}$
with
\begin{eqnarray}
  {\scal c}_n^{\,(N,\pm)} &=& \sqrt{2}\,\sum_m\lb 
    \kappa_m^\pm   \left[S^N\right]_{n,m} 
  +(\kappa_m^\pm)^*\left[S^N\right]_{m,n}\rb\,
  \nonumber\\
  &&+\:\sum_{m,m'}
  (\kappa_m^\pm)^*\,\kappa_{m'}^{\pm}\,\left[S^N\right]_{m,m'} \ .
\label{shift:corrections}\end{eqnarray}
We see that $\langle\bbox{n}|S^N|\pm\bbox{n}\rangle$ can be replaced
by $\langle n|S^N|\pm n\rangle$ at the price of corrections 
${\cal O}(\kappa)$ at most. However, since the left hand side of
(\ref{shift:iterates2}) is of size $\sin(N\Delta\theta_n/2)\sim {\cal
  O}(1)$ (considering the positive sign) for large $N$, the first term
on the right 
hand side must also grow with $N$ and become of ${\cal O}(1)$. In
particular, for $N\gg\kappa/|\Delta\theta_n^{\,(R)}|$, the
$\kappa$-corrections can be neglected. The same holds for the negative
sign in (\ref{shift:iterates2}), but then $N\gg\kappa/|\delta\theta_n|$
must hold. Therefore, whenever either of these lower bounds on $N$
holds simultaneously with the upper bound (\ref{shift:split:con}) we
can write  
\begin{eqnarray}
  \Delta\theta_n^{\,(R)} &\approx& \frac{1}{N}\,\IM
  \Big\{
  \ee^{-iN\theta_n^{\,(0)}}\left[S^N\right]_{n,n}
  \Big\}\ ,\label{shift:shiftSN}\\
  \delta\theta_n &\approx& \frac{2}{N}\,\IM
  \Big\{
     \ee^{-iN\theta_n^{\,(0)}}\left[S^N\right]_{n,-n}
     \Big\}\ .
\label{shift:splittingSN}\end{eqnarray} 
We note that the conditions on $N$ necessary for
Eq.~(\ref{shift:shiftSN}) are met if
$|\Delta\theta_n|>|\delta\theta_n|$ which, as we will find later, is
always the case. However, the conditions for
Eq.~(\ref{shift:splittingSN})  might not be met simultaneously. In this
case, one has to expand in (\ref{shift:alternative}) to obtain 
\begin{eqnarray}
  |\delta\theta_n|\approx\frac{2}{N}\,
\left|\left[S^N\right]_{n,-n}\right|\ , \qquad
\label{shift:altern2}\end{eqnarray}
requiring $\kappa/|\delta\theta_n|\ll N \ll |\delta\theta_n|^{-1}$
--- a condition that can always be fulfilled (if $\kappa\ll 1$).  
However, we will in the sequel calculate the splitting by use of
Eq.~(\ref{shift:splittingSN}), as this expression constitutes a 
linear relation between the splitting and the different
contributions. We therefore assume that the use of
Eq.~(\ref{shift:splittingSN}) is justified and only comment on the use
of Eq.~(\ref{shift:altern2}). (We note that Eq.~(\ref{shift:altern2})
was employed in an earlier account of this work \cite{Doron95b}.)

Returning to Eqs.~(\ref{shift:shiftSN},\ref{shift:splittingSN}) 
recall that any matrix element $[S^N]_{n,m}$ can be expressed as
a {\em sum over paths\/} in matrix element space of length $N$ that
start at $n$ and end at $m$ by writing out the intermediate matrix
multiplications. This yields the real part
of the shift and the splitting as
\begin{eqnarray}
   \Delta\theta_n^{\,(R)} &\approx& \frac{1}{N} \IM\Big\{
   \ee^{-iN\theta_n^{\,(0)}}    {\sum_{\{n\:\rightarrow
    n\}}} \prod_{i=1}^{N-1}\:S_{\lambda_{i},\lambda_{i+1}} \Big\} \ ,  
  \label{shift:shift}\\
  \delta\theta_n\; &\approx&  \frac{2}{N} \IM\Big\{ \ee^{-iN\theta_n^{\,(0)}}
  {\sum_{\{n\:\rightarrow
     -n\}}} \prod_{i=1}^{N-1}\:S_{\lambda_{i},\lambda_{i+1}}\Big\} \ .
  \label{shift:splitting}
\end{eqnarray}
The eigenphase shift and the splitting 
are therefore given in terms of paths of length $N$ that lead 
from index $n$ back to $n$ or to $-n$, respectively.
Note that in order to contribute to the shift, the path must leave the
index $n$ at least once; the trivial path of constant matrix index $n$
does not contribute, as $\exp(-iN\theta_n^{\,(0)})\:[S^N]_{n,n}=1$ is real. 

For the sake of completeness, we also comment on the imaginary
part of the shift. Since $\theta_n^{\,(0)}+\Delta\theta_n$ is real,
$\Delta\theta_n^{\,(I)}=-\IM \{\theta_n^{\,(0)}\}$. 
By unitarity of $S$, one finds
\begin{eqnarray}
  \Delta\theta_n^{\,(I)} \approx -\frac{1}{2}\sum_{m\neq n} 
  |S_{n,m}|^2\ .
\label{shift:imagshift}\end{eqnarray}

It remains to connect eigenphase shifts and splittings to the
corresponding energy quantities. At a given energy $E$ --- which
need not necessarily be an eigenenergy of the system --- the
eigenphases $\theta_j(E)$ are distributed on the unit circle, with the
position of regular doublets determined by
Eqs.~(\ref{shift:phase},\,\ref{shift:shift},\,\ref{shift:splitting}).
Regular doublets revolve around the unit circle with ``velocity''
$\partial\theta_n^{\pm}/\partial E
\approx\partial\theta_n^{\,(0)}/\partial E$ given by the energy
dependence of the $\theta_n^{\,(0)}(E)$.  These quantities can be
taken constant on the scale of the energy splittings, and eigenenergy
splittings are therefore trivially related to eigenphase splitting.  
Consequently, we will in the sequel consider
shifts and splittings of eigen{\em phases\/} rather than
eigenenergies. This has the numerical advantage that we can consider
$S(E)$ at any $E$ without having to worry about quantization.

In order to extract the shift or the splitting of the doublet peaked
at indices $\pm n$ we therefore have to consider paths in $S$-matrix
index space that lead from $n$ back to $n$ or to $-n$,
respectively. Let us briefly focus on the splitting. In principle,
there are two types of paths: {\em direct\/} paths and {\em
chaos-assisted\/} ones \cite{LinBallentine,Bohigas93,Bohigas93a}.  Direct
paths tunnel directly from $n$ to $-n$ in a single transition over a
long distance in phase space. Their contribution to the splitting is
of the order $|S_{n,-n}|$.  Chaos-assisted paths include at least two
tunneling transitions over relatively small phase space
distances. They tunnel from $n$ to some index $\ell$ such that $\ell$
lies in the inner block of classically chaotic motion, then propagate
--- via classically allowed transitions --- to some $\ell'$ within the
inner block and finally tunnel from $\ell'$ to $n$. Contributions
arising from chaos-assisted paths are then of the order $|S_{n,\ell}
\,S_{\ell',-n}|$.  As we have seen in the example of the annular
billiard, tunneling matrix elements $S_{n,\ell}$ fall off very rapidly
(faster than exponential) for large phase space distances
$|n-\ell|$. This rapid decay strongly 
suppresses the contributions from direct paths and explains why the
combination of two tunneling transitions can be much more
advantageous.

\subsection{A Block-Matrix Model}

We now formulate a generalization of the block matrix models
usually encountered in the treatment of chaos-assisted tunneling
\cite{Bohigas93,Tomsovic94} that takes into account the effect
of the transition region between classically regular and chaotic
motion.

Statistical modeling of chaos-assisted tunneling is usually done in
terms of a block matrix model of the type {\em
regular-chaotic-regular} in which properties of the chaotic block are
approximated by random matrix ensembles \cite{Bohigas93}. This
three-block approximation, however, discards all information about
phase space structures inside the chaotic sea, such as the inhibition
of mixing by broken invariant tori (cantori)
\cite{MacKay8484}. Inhibition of classical
transport can lead to dynamical localization of states in regions of
phase space. Chaotic states then do not extend over the full chaotic
region of phase space any more, but only over components of it.
This additional structure would not be reproduced by the approximation
by a single random matrix block.  Block matrix models for
chaos-assisted tunneling were amended to the presence of imperfect
layers in Refs.~\cite{Tomsovic94,Leyvraz96} by introduction of
separate, weakly coupling blocks for each of the phase space
components.  A relationship between classical flux crossing imperfect
transport barriers and quantum Hamiltonian matrix elements was given
in Ref.~\cite{Bohigas90a}.

We now argue that the treatment of chaos-assisted tunneling in a
generic mixed system usually requires a {\em five\/}-block model at
least. The reason is that classical motion in the ``beach'' 
regions close to regular domain is relatively stable, despite of its
long-time chaotic behavior. In particular, transport in the
direction away from the regular phase space region can be strongly
inhibited.  This dynamical stability leads to the formation of quantum
{\em beach states\/} that have most of their amplitude in the beach
region and little overlap with the chaotic sea. It has been reported on
many occasions that beach states have great similarity to regular
states residing on the adjacent island and that they follow EBK-like
quantization rules \cite{Noid77,Swimm79,Bohigas90,Utermann94}.  Note
that the importance of the beach region is highlighted in
chaos-assisted tunneling processes: as tunneling amplitudes
$S_{n,\ell}$ decay rapidly away from the regular island,
chaos-assisted paths of largest amplitude will typically lead to
indices $\ell$ and $\ell'$ such that the corresponding momenta
$P_2(\ell)$ and $P_2(\ell')$ lie {\em just inside\/} the chaotic sea
on either side, that is, in the beach regions.

In order to take account of the special role of the beach regions (or
``edge'' regions, we will use these two expressions as synonyms), we 
propose to generalize the usual three-block model {\em
regular-chaotic-regular\/} to a five-block model of the type {\em
regular-edge-chaotic-edge-regular}.  (In this work we assume that,
apart from the edge layers, no further transport-inhibiting structures
are present. The existence of further transport-inhibiting structures
inside the chaotic sea will require the addition of further blocks.)

\begin{figure}
\centerline{\psfig{figure=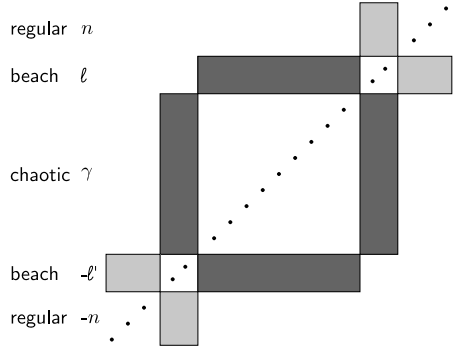,width=6cm}
  \protect\vspace*{.5cm}}
  \caption{The modified block-matrix model:
    structure of blocks participating in the {\em recer\/} contributions.}
  \label{model:fig}
\end{figure}

In the sequel, we approximate $S$ by a five-block model $\tilde{S}$ as
depicted in Fig.~\ref{model:fig} in which each regular region, each
beach region and the center chaotic region are modeled in a separate
block, and coupling between different blocks is weak.  
We assume that $\tilde{S}$ has, by a unitary transformation,
been converted such that all intra-block transitions vanish.  We use
indices $n$ and $-n'$ for the properties of the two regular blocks,
$\ell$ and $-\ell'$ for the beach blocks and $\gamma$ for
chaotic states. For the diagonal elements, we write
$\tilde{S}_{\lambda,\lambda}=\exp(i\tilde{\theta}_{\lambda})$, where
$\lambda\in\{n,\ell,\gamma,-\ell',-n'\}$. Note the symmetries
$\tilde{S}_{n,n}=\tilde{S}_{-n,-n}$ and
$\tilde{S}_{-\ell,-\ell}=\tilde{S}_{\ell,\ell}$, and therefore
$\theta_{-n}=\theta_{n}$,
$\tilde{\theta}_{-\ell}=\tilde{\theta}_{\ell}$.  Inter-block coupling
elements are denoted by $\tilde{S}_{n,\ell}$, $\tilde{S}_{n,\gamma}$,
$\tilde{S}_{\ell,\gamma}$ and so on. It is natural to assume that the
tunneling elements $\tilde{S}_{n,\ell}$ between regular tori and the
beach region will be much smaller than the transition
amplitudes $\tilde{S}_{\ell,\gamma}$ between beach and the center
block.

We now explain how the properties of $\tilde{S}$ can be extracted from
the original matrix $S$. Note that a block that is almost diagonal
$\tilde{S}$ will be changed only little in the transformation to
$S$. The outermost regular blocks are 
diagonal in all orders of $1/k$ (or $\hbar$), and matrix elements of
the transformed matrix will differ from the original ones only by 
exponentially small corrections. We can therefore approximate 
$\tilde{S}_{n,n'}\approx S_{n,n}\,\delta_{n,n'}$. 
The argument also holds for the edge region, since the choice of basis
for the regular blocks will also be good in the edge blocks, and we
can approximate 
$\tilde{S}_{\ell,\ell'}\approx S_{\ell,\ell}\,\delta_{\ell,\ell'}$ and
in the non-diagonal block $\tilde{S}_{n,\ell}\approx S_{n,\ell}$.
Consequently, the choice of the border index between
regular and edge blocks does not affect the results to the present
approximation. 

We model quantum dynamics within the inner block by a superposition of
two Gaussian ensembles \cite{Bohigas84,Blumel88,Blumel90} --- in our
case two Circular Orthogonal Ensembles (COE) \cite{Dyson62a} --- that
approximate the sets of chaotic states with even and odd symmetry,
respectively. For a given regular mode $n$ coupling matrix elements
$\tilde{S}_{n,\gamma}$ are chosen as Gaussian distributed independent
random variables of variance determined by
$\sigma^2_{n,\text{C}}=|\tilde{S}_{n,\text{C}}|^2/(2\ell_{\text{COE}}+1)$
with
\begin{eqnarray} 
  |\tilde{S}_{n,\text{C}}|^2 =
  \sum_{g=-\ell_{\text{COE}}}^{\ell_{\text{COE}}} |S_{n,g}|^2\ ,
  \label{block:effcoup}\end{eqnarray}
where we have taken the chaotic block to extend from
$\ell_{\text{COE}}$ to $-\ell_{\text{COE}}$.  Coupling matrix elements
$\tilde{S}_{\ell,\gamma}$ are defined analogously,
\begin{eqnarray} 
  |\tilde{S}_{\ell,\text{C}}|^2 =
  \sum_{g=-\ell_{\text{COE}}}^{\ell_{\text{COE}}} |S_{\ell,g}|^2\ ,
  \label{block:edge:effcoup}\end{eqnarray}
The choice of $\ell_{\text{COE}}$ contains some
uncertainty that will affect the results as an overall factor in the
effective coupling elements.
Note that an approximation for single coupling matrix elements
$\tilde{S}_{n,\gamma}$, $\tilde{S}_{\ell,\gamma}$ in terms of the
original $S$-matrix elements --- or, even more, as a
semiclassical expression --- is presently not possible, as too little is
known about the precise nature of the quantum localization on the beach
layer. Also, standard semiclassical methods break down in the beach.

For the brevity of notation, we will drop the tildes on $\tilde{S}$
and $\tilde{\theta}$.

\subsection{Extracting the Shift, Splitting and Eigenvector Structure}
\label{block:sum:section}

We now use the block matrix model and formulas
(\ref{shift:shift},\ref{shift:splitting}) to extract approximations
for the shift and the splitting of eigenphase doublets. Also, we
comment on how to extract eigenvector structure and to calculate
eigenphase properties from it.

\subsubsection{Eigenphase Splitting}
When using Eq.~(\ref{shift:splitting}) and the block matrix
representation of $S$ to calculate the splitting, we have to perform the 
sum over paths
\[
{\cal P}^N_{n,-n} \,\equiv\, {\sum_{\{n\:\rightarrow
     -n\}}} \prod_{i=1}^{N-1}\:S_{\lambda_i,\lambda_{i+1}} 
\] 
of length $N\gg\kappa/|\delta\theta_n|$.  As $N$ is large (and under the
assumption that none of the internal diagonal elements
$S_{\lambda_i,\lambda_i}$ is equal to $S_{n,n}$) it is sufficient to
collect contributions that are of order $N$.
We consider three families of paths: paths of type
{\em regular-regular (rr)\/} that contain one single jump from $n$ to
$-n$, paths of type {\em regular-chaotic-regular (rcr)\/} that pass
through the center block, and paths {\em
regular-edge-chaotic-edge-regular (recer)\/} that pass through all
five blocks.  Adding up the different contributions gives the
eigenphase splitting as a sum
\begin{eqnarray}
  \delta\theta_n \approx \delta\theta_n^{(rr)}+
  \delta\theta_n^{(rcr)}+\delta\theta_n^{(recer)}\ .
\label{block:threesplit}\end{eqnarray}
Within each of these families, one has to sum over indices of staying times of
the path at each diagonal element. We show in Appendix \ref{App:splittings}
how this can be done and quote the result: The sum over paths with
$M+1$ jumps passing through $M$ intermediate blocks
$S^{(i_1,i_1)},\ldots,S^{(i_M,i_M)}$ is given as
\begin{eqnarray}
  &&
  {\cal P}_{n,-n}^{N\,(i_1,\ldots,i_M)} \nonumber\\
  &&\sim
  N \ee^{i(N-1)\theta_n^{\,(0)}} \sum_{\lambda_1,\ldots,\lambda_M}
  S_{n,\lambda_1} \prod_{i=1}^{M}
  \frac{S_{\lambda_i,\lambda_{i+1}}}
  {\ee^{i\theta_n^{\,(0)}}-\ee^{i\theta_{\lambda_i}}} \ ,
\label{block:pathsum}\end{eqnarray}
where the sums over the $\lambda_\nu$ run over all indices of the
corresponding blocks $S^{(i_\nu,i_\nu)}$, and $\lambda_{M+1}=-n$.
Corrections to (\ref{block:pathsum}) are of lower order in $N$ or higher
order in transition amplitudes. The phase denominators
\[
 \ee^{i\theta_n^{\,(0)}}-\ee^{i\theta_{\lambda_i}}\:\equiv\:d_{n,\lambda_i} 
\]
arise from the summation over staying times $T_i$ at the different blocks.
Within a given family, each path contributes a factor
$\exp[iT_i(\theta_{\lambda_i}-\theta_n^{\,(0)})]$, and geometrical summation
over $T_i$ results in the denominators listed.  
Note that only phase differences appear that combine the outermost
phase $\theta_n^{\,(0)}$ with one of the phases $\theta_{\lambda_i}$ of the
inner blocks. Contributions containing other phase denominators decay
exponentially in $N$.  
Also, we have only taken into account paths that pass
through each of the inner blocks once. This ``never look back''
approximation is justified since paths containing loops are of higher
order in the inter-block transition elements. For a treatment of loops
in index space, see Appendix \ref{App:splittings}.  
  
Let us now discuss the contributions of the different families in
turn.  For paths of type {\em regular-regular}, we apply
Eq.~(\ref{block:pathsum}) for $M=0$ and find
\begin{mathletters}
\begin{eqnarray}
  \delta\theta_n^{\,(rr)}\:\approx\:
  2\IM\{\ee^{-i\theta_n^{\,(0)}}\,{S_{n,-n}}\} \sim \mid S_{n,-n}\mid\ .
\label{block:split:rr}\end{eqnarray}
Chaos-assisted paths of type {\em regular-chaotic-regular} visit the
center block. By application of Eq.~(\ref{block:pathsum}) for $M=1$,
we find that the {\em rcr\/}-contribution to the splitting is
\begin{eqnarray}
  \delta\theta_n^{\,(rcr)}\:\approx\:
  2\IM\Big\{ \ee^{-i\theta_n^{\,(0)}} \sum_\gamma 
  \frac{S_{n,\gamma}\,S_{\gamma,-n}}{d_{n,\gamma}}
  \Big\} \ .
\label{block:split:rcr}\end{eqnarray}
Finally, the paths {\em regular-edge-chaotic-edge-regular\/} pass
through three intermediate blocks ($M=3$), hence
\begin{equation}
  \delta\theta_n^{\,(recer)}=
  2\IM\Big\{\ee^{-i\theta_n^{\,(0)}}
  \sum_{\gamma,\ell,\ell'}
  \frac{S_{n,\ell}}{d_{n,\ell}}
  \frac{S_{\ell,\gamma}\,S_{\gamma,-\ell'}}
  {d_{n,\gamma}}
  \frac{S_{-\ell',-n}}{d_{n,\ell'}}\Big\}\ .
\label{block:split:recer}\end{equation}
\end{mathletters}
One sees that all tunneling rates mediated by internal blocks can be
strongly enhanced by two effects. 

(1) Combinations of tunneling matrix elements may become
progressively more advantageous as more steps are allowed.

(2) Coherent summation over staying times results in phase
denominators.  Avoided crossings of these phases turn the splittings
into a rapidly fluctuating quantity with respect to small changes in
energy, say, or an external parameter of the system. The phase
denominators also lead to an {\em overall\/} increase in the tunneling
rate since, at a given wave number $k$, there are of order $k$
internal states available. This means that typically, there will be
one phase denominator of size $d_{n,\gamma}^{-1}\sim k/2\pi$ at least.

Both effects, (1) and (2), can also enhance the {\em recer\/}
contributions with respect to the {\em rcr\/} ones. We can therefore
expect the {\em recer\/} contributions to dominate the tunneling rate.
For a given system, their relative importance may vary, depending on
the size of $|S_{n,\gamma}|$ and
$|S_{n,\ell}S_{\ell,\gamma}/d_{n,\ell}|$.

\subsubsection{Eigenphase Shift}

Paths that contribute to the real part of the
shift lead from $n$ back to $n$ and have to leave this index
at least once. To do so, they can either tunnel to the center block
or to the edge block, which leads to a decomposition of contributions
into  
\[
\Delta\theta_n^{\,(R)} \:\approx\: \Delta\theta_n^{(rer)} +
\Delta\theta_n^{(rcr)} +\Delta\theta_n^{(recer)}  \ .
\]
The summation over these paths can be done by the same procedure used
above, and one finds that
the contributions to the real part of the shift are
\begin{mathletters}
\begin{eqnarray}
  \Delta\theta_n^{\,(rer)} &=& \IM\Big\{\ee^{-i\theta_n^{\,(0)}} \sum_\ell
  \frac{S_{n,\ell}\,S_{\ell,n}}{d_{n,\ell}}\Big\}\ ,\!\!\!
\label{block:shift:rer}\\
  \Delta\theta_n^{\,(rcr)} &=& \IM\Big\{\ee^{-i\theta_n^{\,(0)}}\sum_\gamma 
  \frac{S_{n,\gamma}\,S_{\gamma,n}}{d_{n,\gamma}}
  \Big\}\ ,\!\!\!
\label{block:shift:rcr}\\
  \Delta\theta_n^{\,(recer)} &=& \IM\Big\{\ee^{-i\theta_n^{\,(0)}}
  \sum_{\gamma,\ell,\ell'} \frac{S_{n,\ell}\,S_{\ell,\gamma}\,
                    S_{\gamma,\ell'}\,S_{\ell',n}}
                  {d_{n,\ell}\,d_{n,\gamma}\,d_{n,\ell'}}
  \Big\}\ . \!\!\!
\label{block:shift:recer}\end{eqnarray}
\end{mathletters}
Due to the rapid decay of tunneling matrix elements we expect that
$|\Delta\theta_n^{\,(rer)}|\gg
|\Delta\theta_n^{\,(recer)}|\gtrsim|\Delta\theta_n^{\,(rcr)}|$.  
Furthermore from
Eqs.~(\ref{block:split:recer},\ref{block:shift:recer}),
$|\Delta\theta_n^{\,(recer)}|\simeq|\delta\theta_n^{\,(recer)}|$.
Consequently, the shift will typically be much larger than the
splitting.

\subsubsection{Eigenvectors}

We can also use the block matrix model to approximate regular
eigenvector doublets $\bbox{\alpha}^\pm$. We set
$\alpha^\pm_n=1/\sqrt{2}$ and $\alpha^{\pm}_{-n}=\pm1/\sqrt{2}$, and
solve 
\[
  (S-\ee^{i\theta_n^\pm})\,\bbox{\alpha}^\pm=0
\]
to leading order in the coupling matrix elements between neighboring
blocks (neglecting all other coupling matrix elements). We find that
the components of $\bbox{\alpha}^\pm$ in the beach regions are
\begin{mathletters}
\begin{eqnarray}
  \alpha_\ell^{\pm}\approx \frac{1}{\sqrt{2}}\,
  \frac{S_{n,\ell}}{d_{n,\ell}}\;
\label{block:vectors:edge}\end{eqnarray}
and $\alpha^\pm_{-\ell}=\pm\alpha^\pm_\ell$.
Components in the center block are given by 
\begin{equation}
  \alpha_\gamma^\pm=
  \frac{\sqrt{2}}{d_{n,\gamma}}
  \sum_{\ell}
  \frac{S_{n,\ell}\,S_{\ell,\gamma}} {d_{n,\ell}} =
  \sigma_\gamma \sum_{\ell'}
  \frac{\sqrt{2}}{d_{n,\gamma}}
  \frac{S_{\gamma,-\ell'}\,S_{-\ell',-n}} {d_{n,\ell}}\ ,
\label{block:vectors:coe}\end{equation}
\label{block:vectors}\end{mathletters}
if the symmetry $\sigma_\gamma\in\{\pm 1\}$ of the block-diagonalizing
vector $|\gamma\rangle$ is the same as that of $\bbox{\alpha}^\pm$,
and $\alpha_\gamma^\pm=0$ otherwise. In both
Eqs.~(\ref{block:vectors}) the relative error is ${\cal
O}(|S_{n,\ell}|^2/d_{n,\ell}, |S_{\ell,\gamma}|^2/d_{n,\gamma})$.

This approximation for the eigenvectors can be used to relate
eigenphase properties to the size of eigenvector components in the
different blocks. Upon comparison of Eqs.~(\ref{block:shift:rer}) and
(\ref{block:vectors:edge}), we see that the dominant contribution to
the shift can be written as
\begin{eqnarray}
  \Delta\theta_n^{\,(rer)}= 2 \IM\Big\{ \sum_\ell 
  \ee^{-i\theta_n^{\,(0)}}\,d_{n,\ell}\: (\alpha_\ell)^2\Big\}\ ,
\end{eqnarray}
where $\bbox{\alpha}$ is either of the $\bbox{\alpha}^\pm$. Therefore,
the eigenphase shift is related to the eigenvector's overlap with the
beach region.
Similarly, from Eqs.~(\ref{block:split:recer}) and
(\ref{block:vectors:coe}),
\begin{eqnarray}
  \delta\theta_n^{\,(recer)} = \IM \Big\{
  \sum_\gamma 
  \ee^{-i\theta_n^{\,(0)}}\,d_{n,\gamma}\:
  [(\alpha^+_\gamma)^2-(\alpha^-_\gamma)^2]
  \Big\}\ .
\label{block:splitvec}\end{eqnarray}
Eq.~(\ref{block:splitvec}) relates the presumably dominant
contribution to the splitting to the eigenvector's overlap with the
center block.  This explains and quantifies the observation of
Utermann {\em et al.}~\cite{Utermann94}, that regular doublet
splittings in a mixed system are in close correlation with the states'
projection onto the chaotic sea.  Similarly, Gerwinski and
Seba~\cite{Gerwinski94} related tunneling rates between a chaotic
phase space region and a regular island to the overlap of a chaotic
scattering state with the regular island.  However, we see that the
{\em ad hoc\/} association
\begin{eqnarray}
   |\,\Delta\theta_n| \sim \sum_\ell \,|\,\alpha_\ell\,|^2\ ,\qquad
   |\,\delta\theta_n| \sim \sum_\gamma |\,\alpha_\gamma\,|^2
\label{block:adhoc}\end{eqnarray}
is not complete: in the exact relation (\ref{block:splitvec}),
each summand is weighted by a phase difference. 

\subsubsection{Comments}

In view of the explicit formulas, we see that the results are not
significantly changed by approximating the regular and edge blocks of
$\tilde{S}$ by the corresponding elements of the
original matrix $S$.  Only in the immediate vicinity of resonances
between regular and beach eigenphases the approximation
$\tilde{S}_{\ell,\ell}\approx S_{\ell,\ell}$ is not appropriate, as it
over-estimates the imaginary part of the phase $\theta_\ell$ and
leads to a spuriously broad resonance. (With the neglect of
$\Delta\theta_\ell^{\,(R)}$ we shall not be concerned, because we do
not aim at an exact reproduction of the peak positions.) For most beach
states, it is therefore more appropriate to make the somewhat {\em ad
  hoc\/} approximation of $\theta_\ell=\arg\{S_{\ell,\ell}\}$ instead
of $-i\log S_{\ell,\ell}$. This approximation now leads to an
under-estimate of the resonance width, but we have checked that in the
cases discussed in Section \ref{numres:sect} the edge $\ell$ are sufficient
large ($\ell\ge 59$) that the resonances
$|S_{n,n}-\tilde{S}_{\ell,\ell}|^{-1}$ and
$|S_{n,n}-\exp(i\theta_\ell)|^{-1}$ cannot be distinguished by eye.  

\newpage
{}From a methodological point of view, it is worthwhile mentioning
that the formulas for the shift and the splitting can also be
derived from a complementary approach. One can expand the
characteristic polynomial
\[
P_S(x)=\det(S-x)
\] 
of $S$ around $x_n=\exp(i\theta_n^{\,(0)})$ to second order in the external
coupling elements $|S_{n,\lambda_1}|$, $|S_{\lambda_M,-n}|$, 
and then solve for its roots $x^\pm$, $P_S(x^\pm)=0$. Upon definition
of shift and splitting via $x^\pm=x_n+\Delta x \pm \delta x/2$, one
finds formulas that, to lowest order in the internal coupling
elements, are identical to
Eqs.~(\ref{block:split:recer},\,\ref{block:shift:rer}). Hence, the
``never look back'' summation over paths corresponds to the
lowest order of a formal expansion of the eigenvalues, containing
the coupling elements as small parameters.

It is important to realize that the different phase denominators
$d_{n,\ell}$ and $d_{n,\gamma}$ may {\em fluctuate on different
  scales\/} as functions of the energy or an external
parameter. Typically, the center block will be much larger than the
beach blocks, and avoided crossings of $\theta_n^{\,(0)}$ with one of
the $\theta_\gamma$ will occur more often than those with one of the
$\theta_\ell$.  Also, since beach states display EBK-like behavior
with actions that can be similar to those of the regular states,
we can expect phase differences $\theta_n^{\,(0)}-\theta_\ell$ to vary
more slowly than phase differences $\theta_n^{\,(0)}-\theta_\gamma$.  
(For the case of the annular billiard, a semiclassical argument is
given in Ref.~\cite{Frischat97}.)
Consequently, eigenphase splittings $\delta\theta_n^{\,(recer)}$ will show
fluctuations on {\em two\/} scales: There will be a rapid sequence of
peaks due to avoided crossings of regular eigenphases with
chaotic ones and a slow modulation due to the relative motion of regular
and beach eigenphases. 

Let us conclude by summarizing those predictions that genuinely depend
on the explicit inclusion of the beach layers into the five-block
matrix model: 

(I) Eigenphase splitting: Contributing paths typically pass through
all blocks. As a function of an external parameter, the splitting 
varies on two scales: a slow one attributed to the change of the
$d_{n,\ell}^{-2}$, and a rapid one attributed to the change of the
$d_{n,\gamma}^{-1}$. Consequently, one sees resonances of different
line shapes.

(II) Eigenphase shift: Paths contributing to the shift typically visit
only the beach layer. The shift is much larger than the splitting,
and it varies with $d_{n,\ell}^{-1}$ on the slow scale only.

None of these statements would hold for the three-block model, and
therefore (I) and (II) can serve as a test of our five-block model.

\newpage
\subsection{Numerical Results}
\label{numres:sect}

We now give numerical examples of the formulas
just presented. In particular, we will give the most quantitative and direct
proof of the chaos-assisted tunneling picture yet. Also, we will
test the predictions derived from the explicit treatment of the beach
layer in our five-block matrix model. Again, we consider the annular
billiard at parameter values $k=100$, $a=0.4$ and $\delta=0.2$. 

First of all, we have to decide on where to set the borders between
the different blocks. As was already mentioned, the outer borders between
the regular blocks and the edge blocks do not pose any problem
as in both blocks, matrix elements are approximated by the
corresponding matrix elements of the original $S$-matrix.
Due to the tunneling ridges in the region $\ell \gtrsim
k(a+\delta)$, paths starting from high $n$ will most likely tunnel
into this region first. In order to include these paths, we extend
the beach region into the regular block whenever necessary.

The border between the beach region and the chaotic block is more
difficult to determine. In Section \ref{Wigner} we give
numerical evidence that some of the beach states' 
structure arises from 
trapping of classical motion near KAM-like regular island extending
into the chaotic sea down to impact parameters $|L|\approx 0.55$. 
This would suggest the
choice of $|\ell|=55$ for the borders between the edges and the chaotic
block. However, chaotic states that can
carry transport between positive and negative angular momenta have 
sizeable overlap only with angular momentum components between
$\ell=-50$ and $\ell=50$. Therefore, we will take the chaotic block to
extend over 
\newline
\widetext
\begin{figure}
\protect\vspace*{-.4cm}
\centerline{\psfig{figure=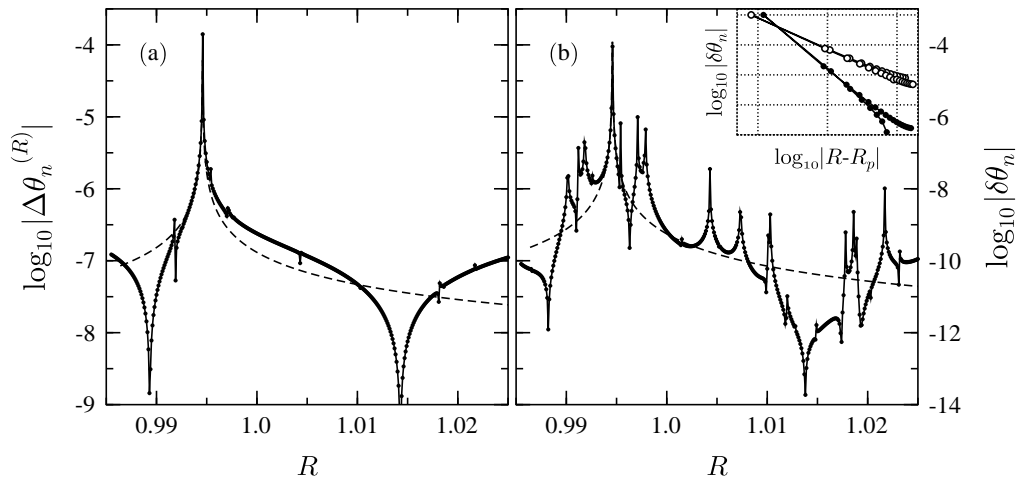,width=15cm}}
  \protect\caption{(a) Eigenphase shift $\Delta\theta_{n}^{\,(R)}$ and
    (b) splitting $\delta\theta_n$ of the doublet $n=70$ for different
    values of the radius $R$ of the outer circle: exact results as
    obtained by numerical diagonalization (circles) and contribution
    predicted by the block matrix model that is dominant between
    $R\approx0.99$ and $R\approx1.015$ (dashed line, see text). 
    Inset: power law dependence of splitting near
    $R_{p}=0.9946$ (full circles) and $R_p=1.0043$ (empty circles).
    Throughout, full lines are to guide the eye.}
  \label{block:splitfig1}
\end{figure}
\narrowtext
\noindent
angular momenta $|\ell|\le\ell_{\text{COE}} =50$.  
The uncertainty inherent in the choice of $\ell_{\text{COE}}$ 
affects the final results via the effective couplings
$S_{\ell,\gamma}$, $S_{n,\gamma}$ which can therefore only be
determined up to an overall constant of order one. 

Let us briefly discuss the magnitudes of the tunneling
amplitudes involved in the different contributions to the splitting.
As expected, the direct  $n$ to $-n$ tunneling matrix element is of 
negligible size. At the present parameter values 
$\delta\theta_{70}^{\,(rr)}\sim|S_{70,-70}|\sim 10^{-60}$ which 
is by many orders of magnitude smaller than the observed splitting
$\delta\theta_{70}\sim 10^{-10}$. 
The difference between contributions $\delta\theta_n^{\,(rcr)}$ and
$\delta\theta_n^{\,(recer)}$ is less drastic. At the parameter values
considered, effective coupling elements
$S_{n,\ell}S_{\ell,\gamma}/d_{n,\gamma}$ are usually by at least an
order of magnitude larger than the corresponding $S_{n,\gamma}$.

The dominance of {\em recer\/} contributions is particularly clear
when studying the behavior of shift and splitting as a function of an
external parameter. We show in Fig.~\ref{block:splitfig1} the shift
and the splitting of the doublet $n=70$ as obtained from numerical
diagonalization of the $S$-matrix as a function of the outer circle's
radius $R=0.985$--$1.025$. $R$ varies over a
sufficiently small interval as to leave the 
classical billiard
dynamics essentially unchanged.  The choice of $R$ as external
parameter has the 
advantage that the tunneling magnitudes given by the {\em inner\/}
scattering matrix $|S_{n,\ell}|=|\SI_{n,\ell}|$ remain constant, as
variation of $R$ affects only the {\em outer\/} scattering matrix
$\SO$.

Fig.~\ref{block:splitfig1}(a) displays the real part of the shift
$\Delta\theta_n^{\,(R)}$. The shift varies slowly as a function of $R$ and 
is over long ranges well reproduced by just a single term of
Eq.~(\ref{block:split:recer}). (The dashed line shows
$|S_{70,59}^2/d_{70,59}|$ with\linebreak
\newpage
\noindent $d_{70,59}\approx\theta_{70}^{\,(0)}-\theta_{59}$  extracted
from $S$, 
slightly shifted to account for $\Delta\theta_{59}$.)
Over relatively large ranges of $R$, one single $\ell$ is clearly
dominant. Transitions between domains of different dominant $\ell$ are
marked by very small shifts due to cancelations between contributing
paths. 
Fig.~\ref{block:splitfig1}(b) shows the splitting of the same
doublet. Note that the logarithmic scale in plot (b) ranges over twice
the number of orders of magnitude than in (a).  As predicted by the
five-block model, the splitting is much smaller than the shift and
shows variations on two scales. The overall, slow modulation is
determined by the beach resonance $\propto d_{70,59}^{-2}$ and closely
follows the behavior of the shift.  On top of this modulation lies a
rapid sequence of spikes that we attribute to quasi-crossings with
eigenphases of the internal block.

For the $\ell=59$ contribution to the splitting (dashed line), we used
Eq.~(\protect\ref{stat:median}) of the next Section to estimate the
median taken over the properties of the chaotic block and divided out
a factor $\sim 15$ to make the dashed line coalesce with the splitting
{\em away\/} from the $d_{n,\gamma}$-resonances.  Note that the
change of dominant edge index $\ell$ is signaled by a strong
cancelation of tunneling paths.
In the inset, we compare the line shapes of the two types of
resonances by plotting $|\delta\theta_{70}|$ as a function of $|R-R_p|$
in a double logarithmic plot near the ``beach'' peak ($R_p=0.9946$,
full circles) and a ``chaotic'' peak ($R_p=1.0043$, empty
circles). Power laws with exponents $-2$ and $-1$ are obeyed to good
precision, thus confirming the prediction of the five-block model.

It is evident that the predictions of the five-block model serve very well to
explain the data. We stress again that the effects just described
--- different line shapes and fluctuations on two parameter scales ---
genuinely depend on the role of the beach layer in tunneling
processes. They serve as clear fingerprints of the quantum
implications of the presence of an beach layer between phase space
regions of classically regular and chaotic motion. 

Let us however mention that the correspondence between shift and
splitting can be less clear. For large $n$ the shift can show
additional modulations that do not appear in the splitting whenever
there is a degeneracy with an beach mode with large $\ell$
\cite{SteffenThesis}. In the shift, this resonance is weighted with
$|S_{n,\ell}|^2$, which favors $\ell$ near the tunneling ridge. In the
splitting, the resonance's contribution has the weight
$|S_{n,\ell}S_{\ell,\gamma}|^2$ which can become very small if $\ell$
is too large. This does however not contradict the predictions (I) and
(II), it merely means that shift and splitting arise by coupling to
different beach modes.

For later purposes it is important to note that whenever the same beach 
state is dominant in both shift and splitting, the correspondence
between the shift and the slow modulations of the splitting  can be
used to ``unfold'' the splitting data from beach properties. By
Eqs.~(\ref{block:split:recer},\ref{block:shift:rer}), the ratio
\begin{eqnarray}
  \widetilde{\delta\theta}_n =
  \frac{4}{\pi}
  \frac{\delta\theta_n}{[\Delta\theta_n^{\,(R)}]^2} 
  \left|\frac{S_{n,\ell}}{S_{\ell,C}} \right|^2 \sim 
  \frac{1}{{|S_{l,\text{C}}|^2}} \left| \sum_\gamma 
    \frac{S_{\ell,\gamma}\,S_{\gamma,-\ell}}{d_{n,\gamma}} \right|
\label{block:unfolded}\end{eqnarray}  
then contains only properties of the center block 
and can therefore be used to extract its ``bare'' quantities.

\subsection{Evolution of Tunneling Flux on the PSOS}
\label{Wigner}
Let us recall that the scattering matrix $S$ is the quantum analogue
of the classical Poincar\'{e} mapping; it constitutes a time
domain-like propagator in the representation fixed by
Eq.~(\ref{scatt:wavefn}). This makes it possible to study the
evolution of ''wave packets'' --- vectors $\bbox{\alpha}_0$
corresponding to initial conditions localized in phase space --- under
the action of $S$. In particular, it is here of interest to follow the
evolution of a tunneling process in phase space.

The comparison of quantum dynamics and classical phase space can, in
the context of the scattering approach to quantization, conveniently
be done by use of Wigner- and Husimi-like functions of quantum
operators \cite{Klakow96,Frischat97}. Let ${\bcal A}$ be some operator
that, for definiteness, we represent in angular momentum basis. As
explained in detail in Ref.~\cite{Frischat97}, ${\bcal A}$ can be
transformed to a function $\rho^{\text{H}}[{\bcal A}](\gamma,L)$ on
the Poincar\'e cell $\CP$ by first performing a Wigner-transform on
${\bcal A}$ and then smoothing with a minimal-uncertainty wave
packet. One obtains
\begin{eqnarray}
  &&\rho^{\text{H}}[{\bcal A}](\gamma,L) = 
  \sum_{\ell \ell'} {\bcal A}_{\ell,\ell'} \times
        \nonumber\\
        &&\exp\left\{\frac{-\Delta\gamma^2}{2}\left[
      \left(kL-\frac{\ell+\ell'}{2}\right)^2 
        + (\ell-\ell')^2\right]
    -i\gamma(\ell-\ell') \right\} \ ,
\nonumber
\end{eqnarray}
where $(\gamma,L)$ are the coordinates in $\CP$, $k$ is the wave
number, and $\Delta\gamma^2$ is a parameter determining the shape of
the smoothing wave packet. We choose $\Delta\gamma^2=4/k$.
In case that ${\bcal A}=\bbox{\alpha}\cdot\bbox{\alpha}^\dagger$ is
the projector of a (normalized) vector $\alpha$, its transform
$\rho^{\text{H}}_{\bbox{\alpha}}(\gamma,L)$ constitutes a positive
semi-definite, normalized density distribution on the Poincar\'e cell,
the {\em Husimi Poincar\'e Density (HPD)\/}. 
By use of HPDs, it becomes possible to follow the phase space
evolution of a tunneling process from one regular torus to its
counterpart. The ``dynamics'' (in iteration count $N$ as the time
variable) of such a process is visualized by projecting iterates
$\bbox{\omega_N}=S^N\bbox{\omega}_0$ of some initial vector
$\bbox{\omega}_0$ onto $\CP$.

Returning to the annular billiard, we consider a starting vector
$\bbox{\omega}_0$ peaked at high angular momentum $n$ and calculate
the Husimi densities
$\rho^{\text{H}}[\langle\bbox{\omega}_N\cdot\bbox{\omega}_N^\dagger\rangle]$
of averaged autocorrelations
\begin{displaymath} 
  \lg\bbox{\omega}_N\cdot\bbox{\omega}_N^\dagger\rg
  \:=\:\sum_{M=0}^{50} \tilde{\bbox{\omega}}_{N+M}\cdot 
  \tilde{\bbox{\omega}}_{N+M}^\dagger
\end{displaymath}

\newpage\widetext
\begin{figure}
\centerline{\psfig{figure=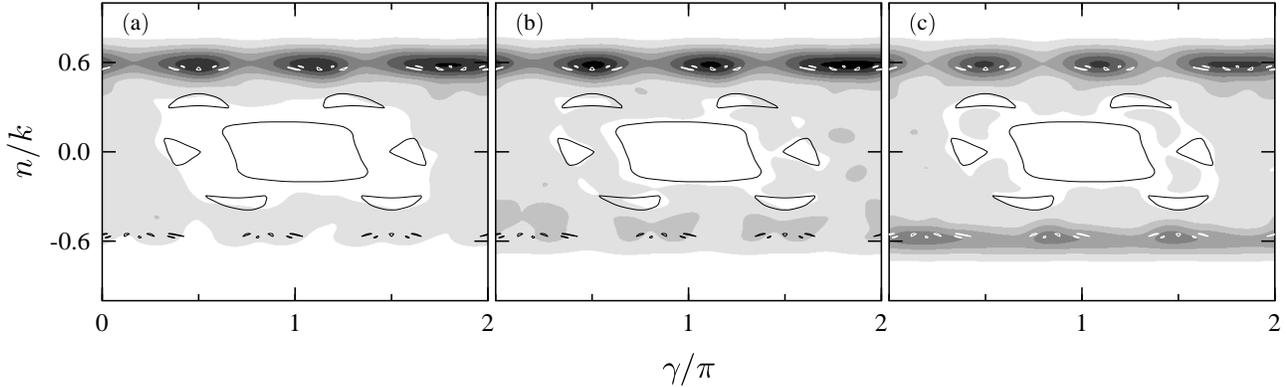,width=18cm}}
\protect\caption{ Spread of tunneling flux over the chaotic sea:
    build-up of quantum amplitude in the beaches. Grayscale plots
    (arbitrary scale) of averaged autocorrelations of the (a) 50-th,
    (b) 100-th, and (c) 500-th iterate of a vector peaked at
    $n=66$. The initial component $n$ has been left out. 
    Full lines indicate classical tori.}
\label{Pdist:dyn:wig1}
\end{figure}
\narrowtext
\noindent
in which the mean over 50 iterations has been performed in order to
average out the internal dynamics of the center block.  The tilde
indicates that the $n$-th component of $\bbox{\omega}$ has been set to
zero. (This truncation is necessary because the smoothing tails of the
large $n$-th component would obscure all features in the
nearby beach regions.)

Fig.~\ref{Pdist:dyn:wig1} depicts the flow of tunneling probability on
$\CP$ by showing
$\rho^{\text{H}}[\langle\bbox{\omega}_N\cdot\bbox{\omega}_N^\dagger\rangle]$
for $n=66$ and $N=50$, $100$ and $500$. At these parameter values, the
tunneling period between WG tori is $2\pi/\delta\theta_{66}\sim
10^7$. As predicted by the five-block matrix model, probability is fed
from the starting angular momentum $n$ into the nearby beach region
until it reaches a value $\sim |S_{n,\ell}|^2$ ($\ell=58$) and spreads
over the chaotic sea up to a value $\sim
|S_{n,\ell}S_{\ell,\text{C}}|^2$, see Fig.~\ref{Pdist:dyn:wig1}(a).
Oscillations between the beaches set in with period
$2\pi/\delta\theta_{58}\sim 1000$, see Fig.~\ref{Pdist:dyn:wig1}(b,c). 
On a much larger time
scale, probability amplitude starts to build up at at $-n$ (not
shown here).  One clearly sees that the shape of the HPD is structured
by the underlying classical dynamics: in the beaches, most
probability builds up around the chains of small KAM-like islands,
whereas in the chaotic sea, the center island and its satellites are
not penetrated. Also, the regions around the satellite
islands and the homoclinic tangles between them are filled only weakly.

\begin{figure}
\centerline{\psfig{figure=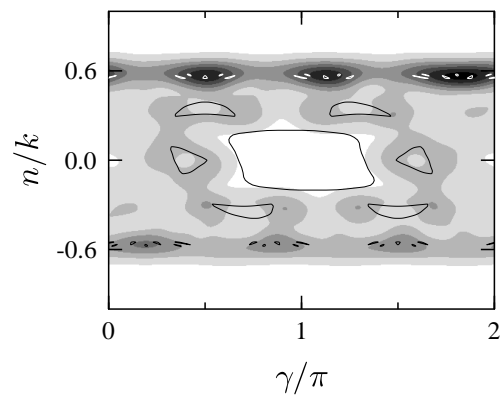,width=18cm}}
\caption{Spread of tunneling flux over the chaotic sea:
    case of a direct degeneracy between the regular doublet and an
    internal state. Grayscale plot (arbitrary scale) of averaged
    autocorrelation of the 4000-th iterate of a vector peaked at
    $n=65$.}  
\label{Pdist:dyn:wig2}
\end{figure}

Chaotic phase space can also be filled in a different
manner, depending on the phases
$\theta_n$, $\theta_\ell$ and $\theta_\gamma$ involved in the
tunneling process.  In Fig.~\ref{Pdist:dyn:wig2} we present the case
of a close degeneracy between $\theta_n$ and
one of the $\theta_\gamma$.  We show
$\langle\bbox{\omega}_{N}\cdot\bbox{\omega}_N^\dagger\rangle$ for
$N=4000$ and the starting vector 
$\bbox{\omega}_0$ peaked at
$n=65$. In this case, there is high probability amplitude in the
sticking regions around the center satellite islands.
\newpage
\vspace*{7.2cm}
\noindent

\begin{figure}
\protect\vspace*{-1cm}
\centerline{
  \psfig{figure=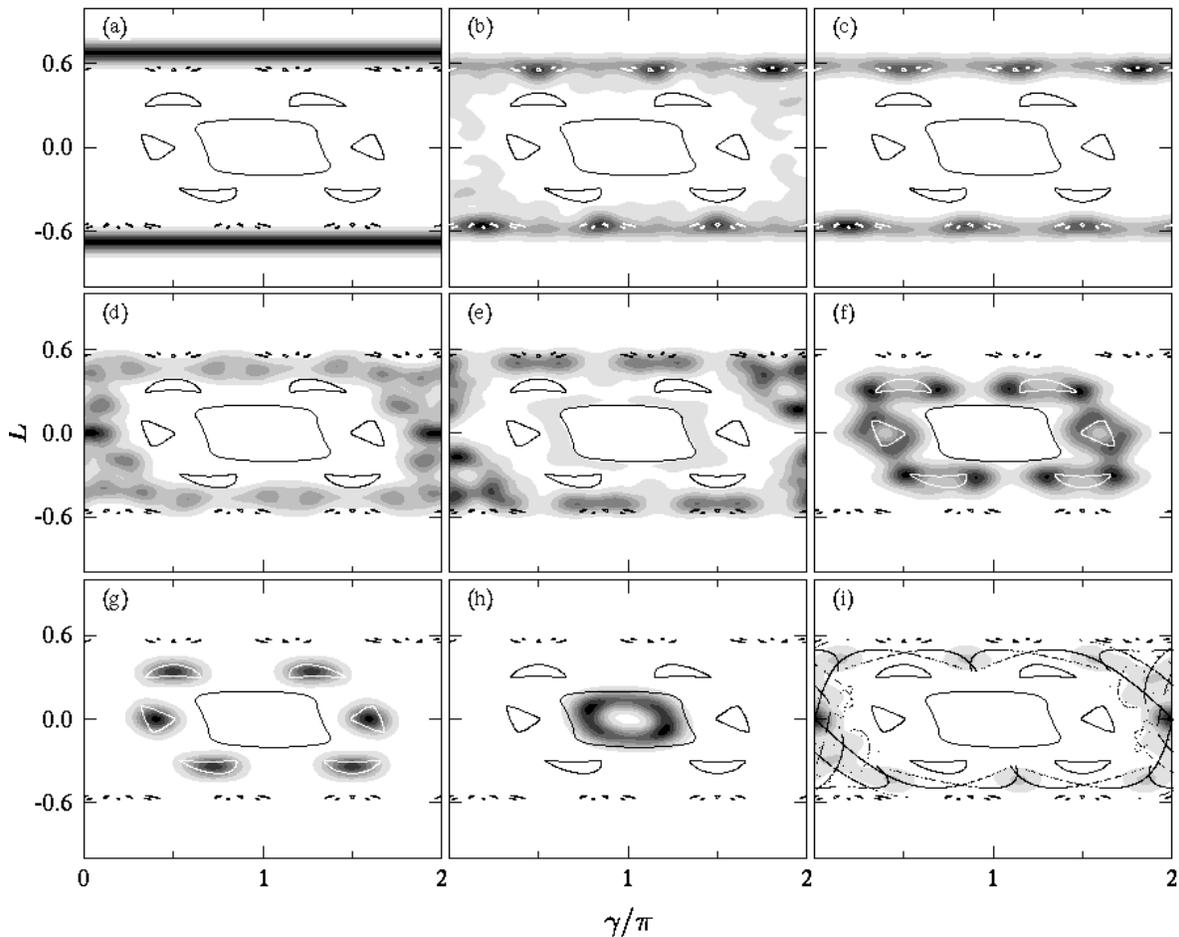,width=16cm}}
\protect\caption{ Poincar\'e Husimi Distributions of
    selected eigenvectors of the annular billiard at $k=100$, $a=0.4$ and
    $\delta=0.2$. (a) Regular high-angular momentum vector, (b,c)
    doublet of beach vectors, (d,e) ``chaotic'' vectors, (f) vector in
    the sticking region around the satellite islands, (g) regular
    vector on the period-6 satellite islands, (h) regular vector
    residing on the main island, and (i) vector scarred by the
    unstable period-1 fixed point and its homoclinic crossings (dots
    depict the stable and unstable manifolds).}  
\label{Pdist:states:fig}
\end{figure}

Finally, we present in Fig.~\ref{Pdist:states:fig} HPDs of nine
eigenvectors of $S$ at $k=100$, $a=0.4$ and $\delta=0.2$ ($k=100$, one
should note, is not an eigenenergy of the annular billiard).  By the
Weyl formula \cite{Baltes} these vectors would, when quantized at a
close-by energy, correspond to the $\sim 2000$-th exited states.  We
show grayscale plots of HPDs $\rho_{\bbox{\alpha}}^{\text{H}}$ with
steps in grayscale
corresponding to equidistant probability contour lines. The overall
scales vary with each sub-plot.  These HPDs at hand, the spread of
tunneling amplitude can now be understood in terms of participating
eigenvectors. For example, the doublet depicted in (b) and (c) is the
beach doublet involved in the tunneling process of
Fig.~\ref{Pdist:dyn:wig1}. Likewise, the vector (f) peaked around the
center satellite islands is the nearly
\linebreak

\newpage
\widetext
\narrowtext
\noindent
degenerate one in the tunneling
process shown in Fig.~\ref{Pdist:dyn:wig2}. Evidently, their shape
forms the spread of tunneling probability on the Poincar\'e cell.

We note that similar figures for much higher wavenumber $k=600$,
corresponding to the $\sim 75000$-th excited states, can be found in
Ref.~\cite{Frischat97}.

\subsection{How Important is Chaoticity?}
\label{dis:chaos:section}

Let us now discuss a numerical study similar to that performed in the
original work by Bohigas {\em et al.\/} \cite{Bohigas93a} in which
regular level splittings are calculated as 
a function of eccentricity $\delta$ at constant $a+\delta$. 
It is important to note that increasing $\delta$ has two effects: 
classical motion in the inner layer $|L|<a+\delta$ becomes chaotic, and
simultaneously tunneling rates from the regular torus to
the chaotic layer are enhanced. {\em A priori\/}, it is not clear
which one of these effects governs the rate at which the splittings
change, but a quantitative estimate of either effect 
has now become possible. \newpage

\begin{figure}
\centerline{
\psfig{figure=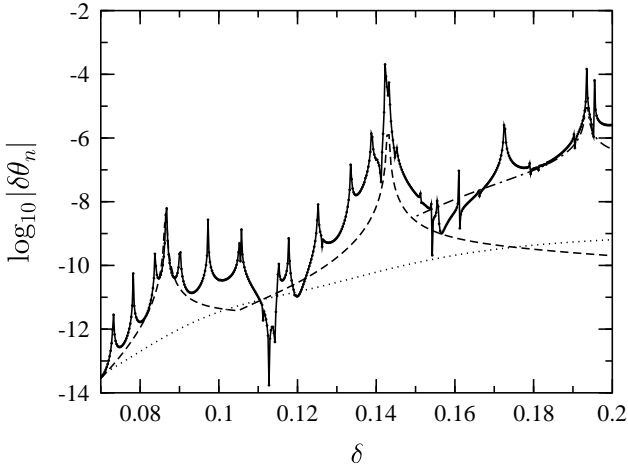,width=10.5cm}}
\caption{Eigenphase splitting $|\,\delta\theta_n|$ of doublet
    $n=39$ as a function of eccentricity $\delta$: exact splittings
    (full line), increase of torus-to-beach tunneling matrix
    elements ($\ell=29$ dotted line, with arbitrary overall factor), 
    and effect of regular-to-beach phase
    denominators (dashed line, with arbitrary overall factor).
    For $\delta>0.15$: tunneling via $\ell=35$ (dashed-dotted line). }
\label{dis:manyd39:fig}
\end{figure}

\vspace*{14.15cm}
We calculated the splittings of high-angular momentum modes for
$\delta=0.03$--$0.2$ and $a+\delta=0.6$ at
$k=60$. Fig.~\ref{dis:manyd39:fig} displays the eigenphase splitting
$|\delta\theta_n|$ for $n=39$  
($n/k=0.65$).  
Exact splittings (full lines) increase over eight
orders of magnitude and roughly follow an exponential increase with
$\delta$. A description of the data in terms of our block-matrix model
that reproduces all the fine details might be a difficult
task --- even with exact $S$-matrix elements at hand --- as the
model relies on classical information to select the
borders between the different blocks and assumes that, apart from the
beach layers, no significant phase space structure is
present. Therefore, the block-matrix model would have to be adjusted
to the varying classical dynamics as $\delta$ changes, and the effect
of remaining structure at lower $\delta$ might have to be taken into
account with the introduction of different blocks. However, the
now-familiar slow modulations in
Fig.~\ref{dis:manyd39:fig} point to the effect of beach layer
states mediating the tunneling flux --- however complicated the
internal structure might be. Indeed, we find that over the range
$\delta=0.07$--0.16 the tunneling processes are mediated by two beach
states peaked around $\ell=29$, with $\ell/k=0.48$ well inside the
non-integrable regime.  In
Fig.~\ref{dis:manyd39:fig}, we have plotted $|S_{n,\ell}|^2$ as a
dotted line (with arbitrary offset) to give a rough estimate of the
change of regular-to-beach tunneling elements via these
particular beach states. 
Taking into account also the effect of beach denominators, we have plotted
$|S_{n,\ell}/d_{n,\ell}|^2$ (dashed line). This estimate is multiplied
with an overall factor $2\cdot 
10^{-5}$ to make the line coincide with the exact data at small
$\delta$. The dashed line already gives a fair
reproduction of the data. It misses only the sharp peaks due to
resonances with the chaotic states, the depression of $\delta\theta_n$
between the $d_{n,\ell}^{-2}$ peaks due to destructive interference,
and the change of coupling strength of the beach state to the chaotic
center states. We conclude that between $\delta=0.07$ and
$\delta=0.2$, the splitting is predominantly determined by the change
of {\em beach\/} parameters, and that the change of internal coupling
between beach and chaotic sea accounts only for a factor of the order
$10$ (difference between the dashed line and the exact data at
$\delta=0.15$).      
For larger $\delta$, different beach doublets take over, but the basic
structure is preserved. The dashed-dotted line displays the {\em
  recer\/}-contributions for $\ell=35$ and $\ell_{\text{COE}}=20$. 

\begin{figure}
\protect\vspace*{-1cm}
\centerline{\protect\hspace*{1cm}
\psfig{figure=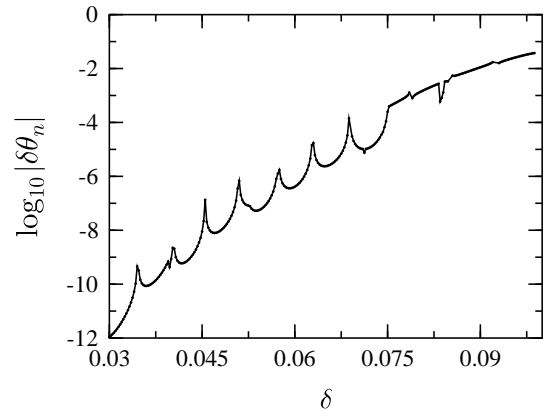,width=10.5cm}}
\caption{Eigenphase splitting of a doublet peaked around
    $\ell=27$ at $k=60$ as a function of $\delta$. Here, resonances arise
    from avoided crossings with states residing on the center island.}
\label{dis:manyd27:fig}
\end{figure}

\begin{figure}
\centerline{
\psfig{figure=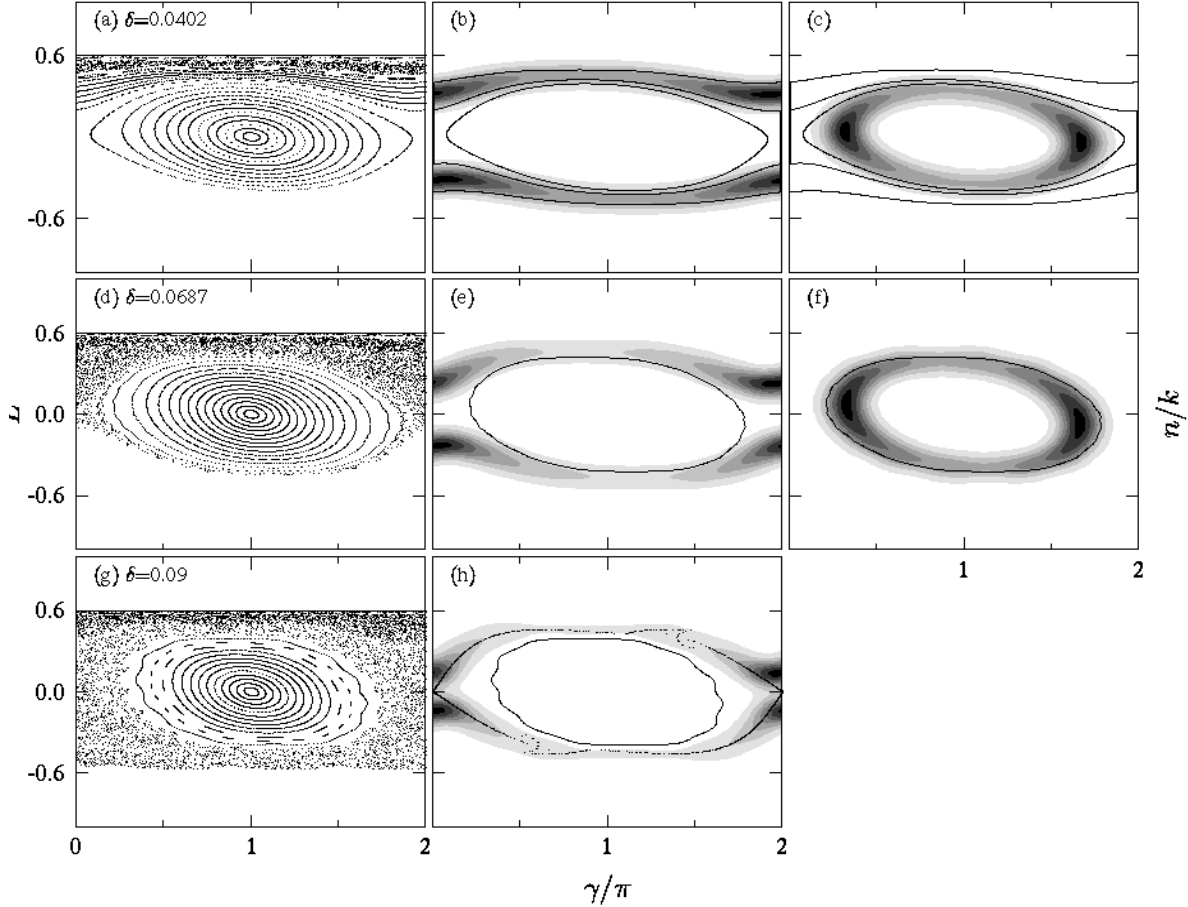,width=16cm}}
\protect\caption{Classical phase space and Husimi Poincar\'e
    distributions of the tunneling doublet of 
    Fig.~\protect\ref{dis:manyd27:fig} and its resonant states at
    (a--c) $\delta=0.0402$, (d--f) $\delta=0.0687$ and (g--h)
    $\delta=0.09$. The doublet resides on KAM-like tori for small
    $\delta$ and dissolves in the chaotic sea for large $\delta$. Note
    that the tunneling at $\delta=0.0687$ takes place between a
    chaotic doublet and is mediated by a regular state. At
    $\delta=0.09$, the dominant tunneling process is direct.}
\label{dis:ratpanel}
\end{figure}

At $\delta$-values below $0.07$, numerical precision does not allow us
to calculate the splitting directly. We can however get an impression
by looking at the splitting of states supported by quasi-integrable
structures at small $\delta$. Presumably, these states will mediate
the tunneling of 
high-angular momentum doublets. In Fig.~\ref{dis:manyd27:fig}, we
show the splitting of the doublet predominantly peaked at $\ell=27$
($\ell/k=0.45$). The splitting shows resonance peaks below
$\delta\approx 0.075$ and then flattens out. This behavior can 
be understood by looking at the HPDs of the states involved. In
Fig.~\ref{dis:ratpanel} we depict  one partner of the tunneling
doublet and its resonant state at (a--c) $\delta=0.0402$, (d--f)
$\delta=0.0687$, and (g--h) $\delta=0.09$. In the corresponding
classical Poincar\'e cells, we have started trajectories from initial
conditions $(\pi,L)$ with $L>0$ only to indicate the classical
inhibition of transport. We can make two interesting observations:
First of all, the resonant tunneling process is mediated by {\em 
regular\/} states residing on the inner island. 
Resonant tunneling via the center island remains the dominant
mechanism, even when classical transport from positive to negative $L$
becomes allowed (d--f). Here, tunneling between a {\em chaotic
  doublet\/} is 
mediated by a {\em regular\/} state. 
Secondly, the outer doublet supported by the KAM-like
tori at small $\delta$ evolves into a doublet of states scarred near
the unstable periodic orbit and stretched along its stable and unstable
manifolds. By quantum localization effects, the doublet structure
persists --- despite of the seemingly chaotic classical motion (see
also \cite{Frischat97}).   
Tunneling between the scarred doublet is direct, as
indications of resonances are absent in the splitting beyond
$\delta=0.075$. 

We are led to the conclusion that the enhancement of tunneling rates
between symmetry-related phase space objects ${\cal A}$ and ${\cal
  T}{\cal A}$ by resonance with quantum states supported by an
intervening phase-space structure ${\cal B}$ is only very loosely
related to the chaoticity of ${\cal B}$, but rather depends
on the {\em topological\/} character of ${\cal B}$. In ${\cal B}$,
it must merely be possible to traverse phase space distance in
\linebreak
\newpage
\widetext
\narrowtext
\noindent
classically allowed steps. It might therefore be more 
appropriate to
refer to ``transport-assisted tunneling,'' a phenomenon of a more
general class than the chaos-assisted tunneling one.

\section{Statistical Analysis of Eigenphase Splittings}
\label{Statistics}

\subsection{Asymptotic Behavior of the Splitting Distribution} 

This Section is devoted to the distribution of level splittings.  We
determine the asymptotic large-$\delta\theta_n$ behavior of the
splitting distribution and find ``typical'' splitting values by
calculating the median of the distribution obtained by extrapolation
of the asymptotic behavior towards smaller splittings.  We assume that
{\em recer\/} contributions dominate and that properties of the beach
blocks vary slowly. It is straightforward to apply the calculation to
the case of {\em rcr\/} contributions as well.\newpage
 
\vspace*{14.35cm}
Starting from Eq.~(\ref{block:split:recer}) we introduce a number of
notational simplifications. For a given $n$, we write phases with
respect to $\theta_n^{\,(0)}$ (that is, set $\theta_n^{\,(0)}=0$),
define $x_\gamma=2\sin(\theta_\gamma/2)$ and collect the coupling of
$\pm n$ to a chaotic state $\gamma$ via the edge into an effective
overlap
\begin{eqnarray}
 && v_{n,\gamma}=\nonumber\\
  && -2\RE\Big\{\,
    \sum_{\ell,\ell'} 
    \ee^{-i(\theta_\gamma+\theta_\ell+\theta_{\ell'})/2}\:
    \frac{S_{n,\ell}\,S_{\ell,\gamma}\,S_{\gamma,-\ell'}\,S_{-\ell',-n}}
       {2\sin(\theta_\ell/2)\,
        2\sin(\theta_{\ell'}/2)}\Big\}\ ,
\end{eqnarray}
which leads to
\begin{displaymath}
  \delta\theta_n^{\,(recer)}\:=\:\sum_{\gamma=1}^{N_\gamma}\:
  \frac{v_{n,\gamma}}{x_\gamma}\ .
\end{displaymath}
$N_\gamma$ denotes the dimension of the center block.
In the sequel we drop the subscript $n$.

To devise a statistical treatment for the center block, we make the
following assumptions concerning the distribution of the
chaotic eigenphases $\bbox{\theta}=\{\theta_\gamma\}$, and the
effective overlaps $\bbox{v}=\{v_\gamma\}$, $\gamma=1,\ldots,N_\gamma$.
We assume that

(1) there is no correlation between the overlaps and the
  eigenphases, 

(2) eigenphases $\theta_\gamma$ are real, ranging from $[-\pi,\pi]$, and 
  the joint distribution function $P(\bbox{\theta})$ of the
  eigenphases either
  (a) is Poissonian, i.e. the eigenphases are uncorrelated, or
  (b) has the property that degeneracies of eigenphases
    $\theta_\gamma$ are suppressed (as is the case in Dyson 
    random matrix ensembles),

(3) the overlaps are mutually independent Gaussian random
  variables with zero mean and variance $\sigma^2$. 
$\sigma$ will be fixed in terms of $S$-matrix elements in the following.
 
The joint probability distribution of the $\theta_\gamma$ and
$v_\gamma$ is then given by 
\[
  P( \bbox{\theta},\bbox{v}) = P(\bbox{\theta})
  \prod_{\gamma=1}^{N_\gamma}\frac{1}{\sqrt{\,2\pi \sigma ^2}}\:
  \exp\left(-\frac{v_\gamma^2}{2\sigma ^2}\right) \ .  
\]
We can write the probability density of $\delta\theta$ in the form 
\begin{eqnarray}
  P(\delta\theta)\: =\:\int \dd\bbox{\theta}\;P(\bbox{\theta})
  \,P( \delta\theta|\,\bbox{\theta}) \ ,
\label{stat:fulldist}\end{eqnarray}
where $P(\delta\theta|\,\bbox{\theta})$ is the conditional probability
of $\delta\theta$ given $\bbox{\theta}$, and the integral is performed
over $[-\pi,\pi]^{N_\gamma}$.  For fixed $\bbox{\theta}$, the
$v_\gamma/x_\gamma$ are mutually independent Gaussian random variables
with variances $(\sigma/\gamma)^2$, and thus
$P(\delta\theta|\,\bbox{\theta})$ is a Gaussian of variance
$\sigma^2\eta(\bbox{\theta})$, where $\eta(\bbox{\theta})$ is given by
\[
  \eta(\bbox{\theta}) =
  \sum_{\gamma=1}^{N_\gamma}\frac 1{x_\gamma^2}\ .
\]
To perform the integration over $\bbox{\theta}$ in
Eq.~(\ref{stat:fulldist}), we introduce 
$\eta(\bbox{\theta})$ as an additional integration variable by
re-writing $P(\delta\theta)$ as
\begin{eqnarray}
  P(\delta\theta)\:=\:
  \int_0^\infty \dd\eta \;
  P(\eta)\,\frac{1}{\sqrt{\,2\pi \sigma ^2\eta }}\:
  \exp\left( -\frac{\delta\theta^{\,2}}{2\sigma^2\eta}\right)   
\label{Pdelta1}\end{eqnarray}
with 
\begin{eqnarray}
  P(\eta)\:=\:\int\dd\bbox{\theta}\: P(\bbox{\theta})\,
  \delta\left(\eta-\sum_\gamma\frac{1}{x^2_\gamma}\right)\ .
\label{Peta:def}\end{eqnarray}
Note that, by virtue of the Central Limit Theorem, almost all overlap
distributions will give rise to a Gaussian form of
$P(\delta\theta|\,\bbox{\theta})$ in Eq.~(\ref{Pdelta1}).

We continue by examining the large-$\eta$ asymptotic behavior of
$P(\eta)$ for the two cases listed in condition (2).
 
(a) If the  $\theta_\gamma$ obey a Poissonian distribution, then $\eta$
is a sum over independent identically distributed variables
$\eta_\gamma$, each of which is distributed as  
\begin{eqnarray}
  P(\eta_\gamma)&=&\frac 1{2\pi}\int_{-\pi}^\pi\dd\theta\;\delta
  \left(\eta_\gamma-\frac{1}{x^2}\right)  
  \nonumber\\
  &=&\left\{ 
  \begin{array}{ll}
    1\,/[\,2\pi\eta_\gamma\sqrt{\eta _\gamma-1/4}\,] 
    & \quad {\rm for}\quad \eta _\gamma\ge 
    \frac 14 \ ,\\ 
    0 & \quad {\rm for}\quad \eta _\gamma<\frac 14\ .
\end{array}
\right.
\nonumber\end{eqnarray}
Note that for  large $\eta_\gamma$ this asymptotically behaves like 
$P(\eta _\gamma)\sim 1/2\pi\eta_\gamma^{3/2}$.
In order to obtain the distribution $P(\eta)$ of the sum, 
we evaluate the characteristic function 
\[
  \hat{P}_1(\omega )=\frac 1{2\pi }\int_{1/4}^\infty
  \dd y\;\frac{\ee^{-i\omega y}}
   {y\sqrt{y-1/4}}=\text{erfc}\left(\frac{\sqrt{i\omega }}{2}\right) \ ,
\]
(see \cite[Eqs.~3.383.4 and 9.236.1]{Gradshteyn}) where
$\text{erfc}(x)$ denotes the complementary error function.
We use the fact that the characteristic function of a sum of $N_\gamma$ 
independent random variables is the product of all their characteristic 
functions, to get
\[
  P(\eta )=\frac 1{2\pi }\int_{-\infty }^\infty\dd\omega \;
  \text{erfc}^{N_\gamma}\left(\frac{\sqrt{i\omega }}{2}\right)
  \:\ee^{i\omega\eta}\ .
\]
The structure of the $\eta\rightarrow\infty$ tail is determined by the
non-analytic behavior of $\hat{P}_1(\omega)$ at the origin. Thus we can
approximate $P(\eta)$ in that regime by expanding
in a power series in $\omega$,  
\[
  \text{erfc}^{N_\gamma}\!\left(\frac{\sqrt{i\omega}}{2}\right) 
  =1-N_\gamma\sqrt{\frac{i\omega }{\pi }}
  +{\cal O}\left(\frac{N_\gamma^2\,\omega}{2\pi}\right)\ . 
\]
The leading (non-analytic) square root term is proportional to $N_\gamma$. 
Thus the frequency dependence scales with $N_\gamma^2$, and the
resulting distribution $P(\eta)$ scales asymptotically as 
$N_\gamma^{-3}\,P(\eta/N_\gamma^2)$. 
We thereby obtain the asymptotic distribution of $\eta$, 
\begin{equation}
  P(\eta)\propto\:\frac {N_\gamma}{\eta ^{3/2}}
  \qquad{\rm for}\quad \eta\rightarrow \infty   \ .
\label{PEtaPoisson}\end{equation}

(b) Next we will evaluate $P(\eta )$ for a general eigenphase
distribution in which the occurrence of eigenphase degeneracies is
suppressed \cite{FelixNote}. In order to re-write the distribution
function Eq.~(\ref{Peta:def}) in terms of products over the
eigenvalues, rather than sums over them, we introduce the integration
parameter $\alpha=\prod_\gamma x_\gamma$ by writing
\[
  P(\eta)\:=\:\int_{-1}^1\!\dd\alpha\!\int\!\dd\bbox{\theta}
  P(\bbox{\theta})\;\delta\!\left(
  \eta-\sum_{\gamma}\frac{1}{x_\gamma^2}\right)  
  \;\delta\!\left( \alpha\!- \!\prod_\gamma x_\gamma \right)
  \ .
\]
In the first $\delta$-function, we can now substitute 
\[
\sum_\gamma\frac{1}{x_\gamma^2}\;=\;\
\frac{1}{\alpha^2}\:\sum_\gamma\,\prod_{\gamma'\neq\gamma}x^2_{\gamma'}
\;\equiv\;\frac{1}{\alpha^2}\,\sum_\gamma\, {\cal X}^2_\gamma\ , 
\]
where we have introduced the notation ${\cal
X}_\gamma=\prod_{\gamma'\neq\gamma} x_{\gamma'}$.  By changing the
integration variable $\alpha\mapsto\alpha/\sqrt{\eta}$ we can extract
the explicit $\eta$-dependence from the first $\delta$-function and arrive
at
\begin{eqnarray}
  P(\eta)=\frac{1}{\eta ^{3/2}} \int \dd\bbox{\theta}
  P(\bbox{\theta}) \int_{-\eta^{1/2} }^{\eta^{1/2}} \!\dd\alpha\: 
  \delta\left( 1-\frac 1{\alpha ^2}
  \sum_\gamma {\cal X}_\gamma^2\right) 
  \nonumber\\
  \times\:\delta\left( \frac{\alpha}{\sqrt{\eta}}-\prod_\gamma
  x_\gamma\right)\ .   
\label{PetaInt}\end{eqnarray}
By inspection of the first $\delta$-function, and recalling that
$|{\cal X}_\gamma|<1$, we see that contributions to the integral can
only arise from the range $|\alpha|<N_\gamma^{1/2}$. However, for
finite $\alpha$, the second $\delta$-function is in the limit
$\eta\ra\infty$ given by
\begin{equation}
  \delta \left( \frac \alpha {\sqrt{\eta }}-\prod_\gamma
  x_\gamma\right) \sim
  \sum_\gamma\,\delta(x_\gamma) \:|{\cal X}_\gamma|^{-1}\ ,
\label{deltaDecomp}\end{equation}
provided all the $x_\gamma$ are {\em distinct\/}. Hence, the second
$\delta$-function is asymptotically independent of $\alpha$,
and the $\alpha$-integration can be performed explicitly, using 
\begin{equation}
  \int_{-\infty}^\infty \dd\alpha \;\delta \left( 1-\frac 1{\alpha ^2}
  \sum_\gamma {\cal X}^2_\gamma\right) 
  =\left(\sum_\gamma{\cal X}^2_\gamma\right)^{1/2}\  .
\label{alphaInt}\end{equation}
Finally, we substitute the form (\ref{deltaDecomp}) of the second
$\delta$-function and the result (\ref{alphaInt}) of the
$\alpha$-integration into Eq.~(\ref{PetaInt}). 
Noting that if $x_\gamma=0$, then ${\cal X}_{\gamma'}=0$ for
all $\gamma'\neq\gamma$, and that $\delta (\theta_\gamma)=\delta(x_\gamma)$, 
we arrive at the distribution 
\begin{equation}
  P(\eta )\sim\frac {N_\gamma}{2\pi \eta ^{3/2}}
  \qquad{\rm for}\quad \eta\rightarrow \infty   \ .
\label{EtaLevi}\end{equation}
It is remarkable that the distributions $P(\eta)$ in
Eqs.~(\ref{PEtaPoisson}) and (\ref{EtaLevi}) display the same
asymptotic power-law dependence.

Finally, we note that, by virtue of the Gaussian form of the integrand
in Eq.~(\ref{Pdelta1}), the integral depends primarily on large
$\eta\ge\delta\theta^{\,2}/\sigma^2$.  To extract the asymptotic
behavior of $P(\delta\theta)$, we only need the asymptotic form
of $P(\eta)$ for large $\eta$. 
Inserting (\ref{EtaLevi}) into Eq.~(\ref{Pdelta1}) gives 
\begin{equation}
  P(\delta\theta)\sim N_\gamma \int_0^\infty \dd\eta \;
  \frac {\ee^{ -\delta\theta^{\,2}/2\eta \sigma^2}}
  {(2\pi)^{3/2}\eta^2\sigma }
  =\frac{\sigma N_\gamma}{\pi \sqrt{2\pi}\,\delta\theta^{\,2}}
\label{PKhatFinal}\end{equation}
for large $\delta\theta$. 
This is the large-$\delta\theta$ tail of a Cauchy distribution, in
accordance to the prediction of Leyvraz and Ullmo \cite{Leyvraz96}. 
We stress that we have derived this result for the case of a
Poissonian eigenphase distribution and, in a second derivation,
without assuming any explicit form of the joint eigenphase
distribution function.
In the latter case, we only had to 
make the assumption that the joint distribution function vanishes 
whenever two eigenphases approach each other.   
Our derivation is therefore more general than the one given in
\cite{Leyvraz96}.

In a broader context, it is also interesting to mention that the
calculation (b) can be generalized to the case of a distribution
\begin{eqnarray}
  P_\nu(\delta\theta)\:=\:\int\dd\bbox{\theta}\;P(\bbox{\theta})\;\delta
  \left( \eta - \sum_\gamma \frac{1}{x_\gamma^{2\nu}}\right)
\label{stat:generaldist}\end{eqnarray}
which corresponds to the distribution resulting from a sum over paths
that contains a \mbox{$\nu$-th} power of the phase denominator.
The procedure is analogous to the case $\nu=1$ just described, but 
for a substitution $\alpha\ra\alpha/\eta^{1/2\nu}$, and one arrives at 
$P_\nu(\eta)\:=\:(2\pi\nu)^{-1}N_\gamma/\,\eta^{-(2\nu+1)/2\nu}$.
This leads to a large-$\delta\theta$ splitting distribution
\begin{eqnarray}
  P(\delta\theta)\propto\frac{1}{ \delta\theta\,^{(\nu+1)/\nu}}\ .
\label{stat:general}\end{eqnarray}

Let us return to Eq.~(\ref{PKhatFinal}).
Having integrated out the eigenphase dependence of $P(\delta\theta)$, 
we are left with the determination of the variance
$\sigma^2$ of the effective overlaps $v_{n,\gamma}$. 
It is given by
\begin{eqnarray}
  &&\sigma ^2\approx 
  \nonumber\\
  &&\left\langle \left| \,2\,\RE \sum_{\ell ,\ell'} 
  \ee^{-i(\theta_\gamma+\theta_\ell+\theta_{\ell'}) /2}\:
  \frac{S_{n,\ell}\,S_{\ell,\gamma}\,S_{\gamma,-\ell'}\,S_{-\ell',-n}}
  {2\sin(\theta_\ell/2)\,
   2\sin(\theta_{\ell'}/2)}
  \,\right| ^2\right\rangle \ .
\nonumber\end{eqnarray}
We assume the phases of the $S_{\ell ,\gamma}$ to be arbitrary and
uncorrelated, assume the $\theta_\ell$ to be real,
and absorb all phases into a random phase factor $\exp (i\phi _{\ell
,\ell',\gamma})$. Using $|S_{\ell ,\gamma}|=|S_{\gamma,\ell
}|=|S_{-\ell ,\gamma}|=|S_{\gamma,-\ell }|$ we can write
\begin{eqnarray}
   \sigma ^2 &\approx& \left\langle \left| \,\sum_{\ell ,\ell'}\,
   \frac{| S_{n,\ell} S_{n,\ell'} S_{\ell,\gamma} S_{\ell',\gamma} |
     \,\cos\phi_{\ell,\ell',\gamma}}
   {2\,\sin(\theta_\ell/2)\,
       \sin[(\theta_{\ell'}-\theta_n^{\,(0)})/2]} \,\right| ^2 
   \right\rangle 
   \nonumber\\[.2cm]
   &=&\:\sum_{\ell,\ell'}\,
   \frac{| S_{n,\ell}S_{n,\ell'}|^2\, 
         \left\langle| S_{\ell,\gamma} S_{\ell',\gamma}
           |^2\right\rangle }
        {8\,\sin^2(\theta_\ell/2)\,\sin^2(\theta _{\ell'}/2)}\ ,
\label{stat:randphas}\end{eqnarray}
using the fact that the $\phi _{\ell ,\ell',\gamma}$ are uncorrelated
and equi-distributed on the interval $[-\pi ,\pi ]$.  Let us now write
$|S_{\ell,\gamma}|^2=N_\gamma^{-1} |S_{\ell,\text{C}}|^2
\xi_{\ell,\gamma}^2$, where
$|S_{\ell,\text{C}}|^2=\sum_{\gamma}|S_{\ell,\gamma}|^2$ is the total
coupling of the $\ell$ state to the chaotic block, see also
Eq.~(\ref{block:effcoup}), and where the $\xi_{\ell,\gamma}$ are
independent Gaussian variables with unit variance. Using that $\langle
\,\xi_{\ell,\gamma}^2 \xi_{\ell',\gamma}^2\,\rangle =(2 +
\delta_{\ell,\ell'})$ we find a statistical enhancement of the
diagonal ($\ell =\ell'$) terms. More importantly,
the sum is dominated by the terms with the smallest phase
denominators. Consequently, we can neglect the non-diagonal terms and
write
\begin{equation}
  \sigma ^2\approx \frac 3{8 N_\gamma^2}\sum_\ell \left| 
  \frac{S_{n,\ell }S_{\ell {\rm C}}}{\sin (\theta _\ell/2)}\right| ^4\ .
\label{stat:variance}\end{equation}

\subsection{Median Splittings}

As we have just shown, the splitting distribution behaves asymptotically 
like $P(\delta\theta)\sim\delta\theta^{\,-2}$, and it is well known
that the mean of a Cauchy distribution does not exist. Therefore, a
``typical'' value for the level splittings must be obtained
otherwise. We propose to consider the {\em median\/}
$|\delta\theta|_M$ of $|\delta\theta|$ defined by 
\[
  2\,\int_{|\delta\theta|_M}^\infty
  \dd(\delta\theta)\;P(\delta\theta) \:=\:\frac{1}{2}\ .
\] 
The factor two on the left hand side arises from the fact that we
integrate over positive $\delta\theta$ only. 
By extrapolation of the asymptotic form of $P(\delta\theta)$ as given
by Eq.~(\ref{PKhatFinal}) towards smaller $|\delta\theta|$, we find
\begin{eqnarray}
  |\delta\theta|_M \sim \frac{4\sigma N_\gamma} {\pi \sqrt{2\pi}}\ .  
\label{KMedian1}\end{eqnarray}
Inserting the variance $\sigma^2$ of Eq.~(\ref{stat:variance}) into 
Eq.~(\ref{KMedian1}) we finally get for the median splitting  
\begin{equation}
  |\delta\theta|_{M,n}
  \:\approx\: \frac 1\pi
  \left(\sum_\ell \left| \frac{S_{n,\ell} S_{\ell,\text{C}}}
    {\sin[(\theta_\ell-\theta_n^{\,(0)})/2]}
  \right|^4\:\right)^{1/2}\ ,
\label{stat:median}\end{equation}
where we have inserted the index for the $n$-dependence again, as well
as the phase $\theta_n^{\,(0)}$. 
Formula (\ref{stat:median}) for the median splittings estimates the
enhancements of tunneling splittings due to chaos-assisted processes
and constitutes one of the central results of this work. Note that all
quantities appearing in Eq.~(\ref{stat:median}) are defined in terms
if the original $S$-matrix, and the most a {\em direct\/} and
{\em quantitative\/} check of the chaos-assisted tunneling picture
yet becomes possible.

We note that Eq.~(\ref{stat:median}) can be used to recast
Eq.~\ref{PKhatFinal}) for the splitting distribution in a numerically
more convenient form
\begin{eqnarray}
  P(\delta\theta_n)\:=\:\frac{\:\:|
    \delta\theta|_{M,n}}{4\,\delta\theta_n^2}\ .
\label{stat:dist:conv}\end{eqnarray}

We now turn to the discussion of the approximations made in the
derivation of the central results
Eqs.~(\ref{PKhatFinal},\ref{stat:median},\ref{stat:dist:conv}) for the
splitting distributions and the median splittings.  There are four
sources of error. (i) The estimate
(\ref{stat:variance},\ref{block:edge:effcoup}) for the variance
$\sigma^2$ is correct only within an order of magnitude due to the
ambiguity of $\ell_{\text{COE}}$ (see discussion after
Eq.~(\ref{block:edge:effcoup})). Up to now, an {\em a priori\/}
determination of $\ell_{\text{COE}}$ is not possible. Note, however,
that the size $N_\gamma$ of the center block does not enter in the
expressions.  (ii) The effect of imaginary parts of the
$\theta_\gamma$ is not included in our calculation. By unitarity of
the block-transformed matrix,
$\IM\{\theta_\gamma\}\sim\sum_\ell|S_{\ell,\gamma}|^2$, which in the
splitting distribution introduces a cutoff of the Cauchy-like tail at
$|\delta\theta|\sim|S_{n,\ell}/\sin(\theta_\ell/2)|^2$. The resulting
relative correction of the median splitting is ${\cal
O}(|S_{\ell,\text{C}}|^2)$. (iii) Extrapolation of the asymptotic tail
towards smaller $\delta\theta$ is another source of error of
${\cal O}(1)$. (For example, the median calculated from an exact
Cauchy distribution $1/\pi(1+x^2)$ is $1+\sqrt{2}\approx 2.4$,
whereas the median estimated by integrating over its tail
$1/\pi x^2$ is $4/\pi\approx 1.3$.)  (iv) Our
five-block model neglects the effect of transport barriers other than
the one separating the beach from the center of the chaotic
block. Further transport barriers lead to the inhibition of
tunneling flux and thereby decrease the splitting.

We conclude that (\ref{stat:median}) reproduces the exact median
splittings only up to a factor of the order one. If the neglect of
remnant phase space structure is the dominant source of error, then
Eq.~(\ref{stat:median}) gives an over-estimate. However, the error
is expected to be independent of $n$, and we can correct for it by
introducing an {\em overall\/} factor $c$ that we extract from the
numerical data. The formula Eq.~(\ref{stat:dist:conv}) for the
splitting distribution function has to be corrected correspondingly.

We finally return to the issue of the two different representations
Eq.~(\ref{shift:splittingSN}) and Eq.~(\ref{shift:altern2}) for the
splitting that differ by taking either imaginary parts of
$\exp(-iN\theta_n^{\,(0)})[S^N]_{n,-n}$ or absolute values of
$[S^N]_{n,-n}$. In a statistical treatment, these two approaches give
slightly different results, because in the average over the random
phases $\phi_{\ell,\ell'\gamma}$, one obtains
$\langle|\ee^{i\phi_{\ell,\ell,\gamma}}|\rangle=1$ after taking
absolute values, as opposed to
$\langle\cos^2(\phi_{\ell,\ell,\gamma})\rangle=1/2$ after taking
imaginary parts. The median splittings derived from
Eq.~(\ref{shift:altern2}) would therefore be twice the splittings
predicted in Eq.~(\ref{stat:median}). This explains why a formula
given by us earlier \cite[Eq.~(8)]{Doron95b} differs by a factor two
from the one in Eq.~(\ref{stat:median}).

\subsection{Numerical Results}
\label{stat:res:section}

This Section is concluded by a presentation of 
numerical data for the eigenphase splitting and its
distribution for the annular billiard.
We choose parameter values $k=100$, $a=0.4$, $\delta=0.2$ and
vary the outer radius $R$ over $490$ values between $R=0.985$ and
$R=0.1035$. Recall that changing $R$ leaves the $|S_{n,\ell}|$
constant and changes only the eigenphase configuration and the
couplings $S_{\ell,\gamma}$. 
\begin{figure}
\protect\vspace*{-.5cm}
\centerline{
\psfig{figure=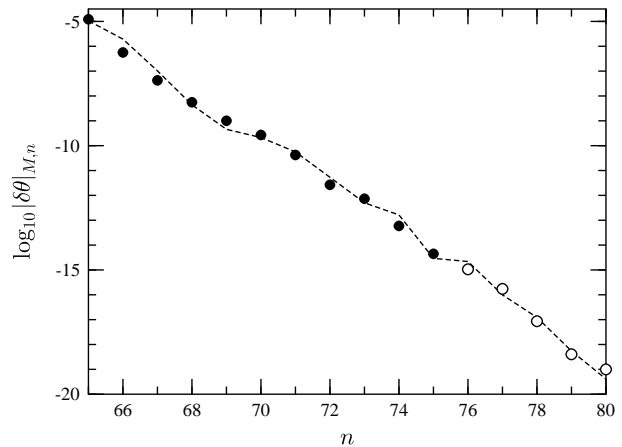,width=11.5cm}}
\protect\vspace*{-.5cm}
\protect\caption{Median splittings $|\delta\theta|_{M,n}$ as a
    function of angular momentum (logarithmic plot). 
    Exact splittings are obtained from numerical diagonalization (full
    circles), calculation of $|[S^N]_{n,-n}|$ for large $N$ (empty
    circles), and estimates are taken from the median formula
    (\protect\ref{stat:median}) and corrected by $c=1/6$ (dashed line). }
\label{stat:angmom:fig}
\end{figure}
Fig.~\ref{stat:angmom:fig} depicts the
median values of eigenphase splittings of doublets peaked at angular
momenta $n=65,\ldots,80$ as a function of $n$. Full circles represent
median splittings $|\delta\theta|_{M,n}$ as obtained from numerical
diagonalization.  Since the diagonalization routine cannot
differentiate between eigenvalues which are closer than $\sim 10^{-15}$,
splittings beyond $n=75$ could not be resolved directly. Instead, they
were extracted by use of Eq.~(\ref{shift:altern2}), that is, by
numerical calculation of  $2[S^N]_{n,-n}/N$ for large
$N\lesssim|\delta\theta_n|^{-1}$.  
Empty circles represent median splittings
thus obtained from $30$ configurations with $R$ ranging from $R=1$ to
$R=1.3$. We have taken the edge region to extend over angular momenta
$\ell=56,\ldots,64$ and have chosen $\ell_{\text{COE}}=50$.  To
account for the over-estimate of the splitting by
Eq.~(\ref{stat:median}), theoretical predictions are multiplied by an
overall factor $c\approx 1/6$.  The dashed line shows the resulting
approximation for the median splittings. Apart from the factor $c$,
the formula (\ref{stat:median}) is in good agreement with the exact
median splittings.

\begin{figure}
\centerline{
\psfig{figure=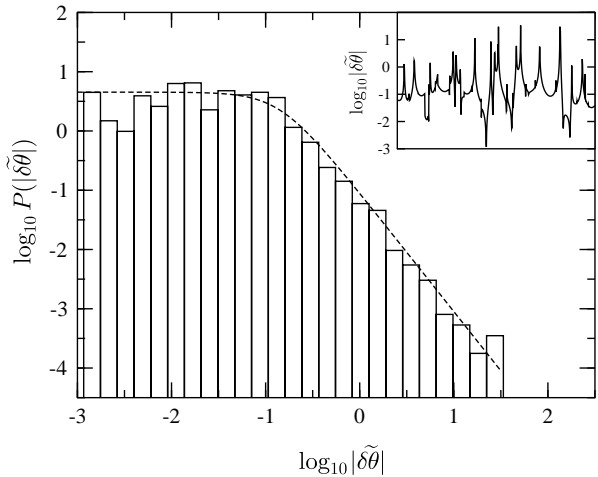,width=11.5cm}}
\caption{Distribution function $P(|\widetilde{\delta\theta}|)$
    for the ``reduced'' splitting $|\widetilde{\delta\theta}|$ (solid line),
    compared to the predicted Cauchy distribution (dashed line),
    double logarithmic plot. Inset: ``reduced'' splittings obtained
    after unfolding the modulations due to the beach layers.}
\label{stat:unfolded:fig}
\end{figure}

Let us turn to the splitting statistics. If one is interested in the
fluctuations due to changes in the chaotic dynamics, one first has to
discard the slow modulation due to the change of beach layer
properties. We do so by considering the ``reduced'' splitting
$\widetilde{\delta\theta}_n$ of Eq.~(\ref{block:unfolded}) for those
values of $R$ at which one single $\ell$ is dominant in both shift and
splitting (for $n=65,\ldots,67$, and here $\ell=n-7$).  We find that
the median of $\widetilde{\delta\theta}_n$ is approximately equal to
$c$, independent of $n$ (not shown).  Fig.~\ref{stat:unfolded:fig}
confirms that the distribution $P(|\widetilde{\delta\theta}|)$ falls
off like a Cauchy distribution of width $c$. For the figure, 750 exact
splittings were transformed to reduced ones (see inset) and collected
in a histogram with log-binning (main figure, solid line). This is
compared to the Cauchy distribution
$2c\pi^{-1}/(c^2+\widetilde{\delta\theta}{}^{\,2})$ (main figure,
dashed line).  The agreement is very good.
 
\begin{figure}
\centerline{
\psfig{figure=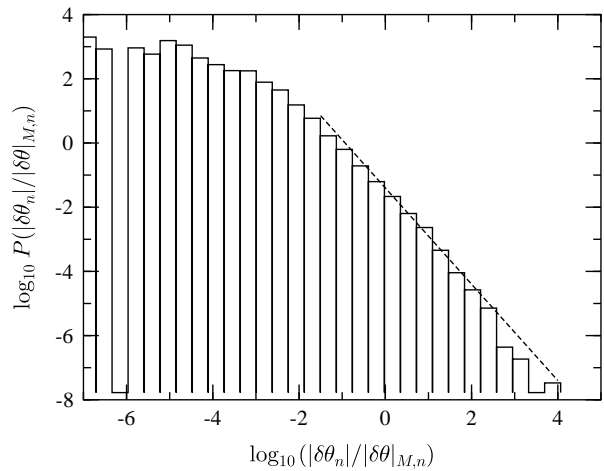,width=11.5cm}}
\caption{Distribution of original eigenphase splittings
    $|\delta\theta_n|$, for each $n$ divided by the median
    $|\delta\theta|_{M,n}$ (solid line), double logarithmic
    plot. Dashed line: comparison to a $\delta\theta^{-3/2}$ power-law
    decay, prefactor fitted to the data.}
\label{stat:folded:fig}
\end{figure}

Finally, we show in Fig.~\ref{stat:folded:fig} the distribution
function $P(|\delta\theta_n|/|\delta\theta|_{M,n}\!)$ of exact
splittings divided by the median splittings displayed in
Fig.~\ref{stat:angmom:fig}. The actual splittings display
a power law $P(|\delta\theta|)\sim |\delta\theta|^{-3/2}$ (dashed
line).  This can be understood by realizing that the variation in $R$
is sufficiently large to average not only over avoided crossings
between chaotic and regular eigenphases, but also over avoided
crossings between regular and beach eigenphases.  These avoided
crossings, however, appear in the sum (\ref{block:split:recer}) with a
squared phase denominator $d_{n,\ell}^{-2}$, and we have argued in
Eq.~(\ref{stat:general}) that the distribution generated by such
contributions displays a power-law decay with exponent $-3/2$.  The
exponent $-3/2$ conforms to the findings of Leyvraz and Ullmo
\cite{Leyvraz96}, who studied chaos-assisted tunneling 
in the presence of a imperfect transport barrier in the chaotic sea. 

\section{Discussion}
\label{disc:sect}

\subsection{Possible Experimental Realizations}

Even though the occurrence of chaos-assisted tunneling should be a
very general phenomenon, an experimental proof of the effect has not
yet been given. The main difficulty might not be to {\em measure\/} the
effect, but to {\em recognize\/} it. As long as little is known about
tunneling in multi-dimensional mixed systems in general, it will be
difficult to separate out the different contributions to
the tunneling rates and to identify the effects of classical
transport. It is our strong suspicion that, as soon as qualitative
theories for experimental systems are developed,
chaos-assisted tunneling will turn out to be a frequent effect in the
splitting of dynamical tunneling doublets.  

\subsubsection{Superconducting Microwave Cavities}

It has been argued in this work that the annular billiard serves as an
excellent paradigm for chaos-assisted tunneling. An experimental
realization of it is presently investigated by the Darmstadt group of
Richter {\em et al.\/} \cite{Hofferbert}.
Performing resonance measurements on a superconducting niobium
microwave cavity, the Darmstadt group has extracted high-quality
spectra in the frequency range 0--20 GHz, corresponding to $k=0$--50 in our
units, but experimental accuracy does not yet allow a resolution
of the splittings of high-angular momentum doublets. However, it might
be just as interesting to measure the energy splittings of {\em beach\/}
doublets, as these
splittings are also chaos-assisted, but by orders of magnitudes larger
than those of the regular doublets. 

\subsubsection{Atomic Systems}

Atomic systems have served as paradigms of many predictions in quantum
chaos, and there are some atoms in which an observation of
chaos-assisted tunneling might be conceivable. Hydrogen in a weak
magnetic field 
\cite{FriedrichDelande} may be such a case. In the weak field
limit, the classical system has symmetry-connected regular islands
corresponding to low-angular momentum motion along the field axis on
either side of the hydrogen core. These islands are separated by a
chaotic sea, and dynamical tunneling between the corresponding
low-angular momentum quantum states should therefore be enhanced by
chaos-assisted processes.  It is however not clear, whether these
splittings are large enough to be experimentally accessible. Another
much-studied system in the field of quantum chaos, the quantum kicked
rotor, has recently been experimentally realized by Moore {\em 
et al.}~\cite{Moore95} using ultracold sodium atoms in pulsed,
near-resonant light.  Rotor systems have been considered by several
groups in studies of dynamical tunneling in the presence of chaos
\cite{Casati94,Averbukh95}, and a link between theory and experiment
might soon become possible. Again, the experiment might still be far
from the required degree of accuracy.

\subsubsection{Open Systems}

Finally, chaos-assisted processes can enhance not only tunneling
oscillations, but also the {\em decay\/} of regular modes in a mixed,
open system in which the dominant coupling to the continuum is
mediated by states residing on the chaotic layer. Experimental
realizations of such systems were studied by N\"ockel {\em et al.\/}
and others \cite{Noeckel9496} who considered the
$Q$-spoiling of 
whispering gallery modes in deformed lasing droplets. In a recent
work, Hackenbroich and N\"ockel \cite{Hackenbroich97} also considered
mixed systems in which the direct coupling of regular modes to the
continuum is suppressed, but where chaotic states have sizable
coupling to the continuum. Regular modes may then decay via a
multi-step process of type {\em regular-chaotic-continuum\/}.
Their results were motivated by a study of a modified
version of the annular billiard, in which the outer circle is replaced
by a mirror, and the billiard is supposed to have higher optical
density that the exterior region outside the mirror. It was found that
chaos-assisted decay can lead a dramatic enhancement of level widths.

\subsection{Discussion and Conclusions}

Having discussed the experimental perspectives, one is immediately led
to the question of the general applicability of the method and the
block-matrix model proposed in this work. 

It is clear that the scattering problem must be solved separately for
each system under consideration. In a general system, it might be very
difficult to formulate the $S$-matrix and to find a basis in which the
$S$-matrix elements $S_{n,n}$ and $S_{-n,-n}$ corresponding to motion
on the tunneling tori are sufficiently close to unity. It must however
be said in favor of the scattering approach that the difficulty of
finding an EBK-quantization scheme in non-separable systems is by no
means inherent to the scattering approach, but presently poses one of
the most serious problems of semiclassical theory in general
\cite{Tomsovic9596}. (In fact, it is one of the fortunate
aspects of the annular billiard that the angular momentum basis is
semiclassically diagonal in the region of regular motion.)

In situations where $S_{\pm n,\pm n}$ are not sufficiently close to
unity, Eqs.~(\ref{shift:shiftSN1},\ref{shift:splittingSN1}) may still
work well, provided that sufficient knowledge of eigenvector structure is
available (see \cite{Hackenbroich97b} for an application to the
case of rough billiards \cite{Frahm97} and other systems). However,
calculation of the splitting by summation over long paths from $n$ to
$-n$ relies on a sufficient localization of $|\bbox{n}\rangle$ at the
$n$-th component.  

Clearly, the block matrix model used in the summation over paths must
be adopted to the specific transport situation encountered. In case of
structure other than the beach layers, additional blocks must be
introduced. This does however not lead to problems in the summation
Eq.~(\ref{block:pathsum}) as long as the outermost tunneling element
is smaller than any of the internal coupling elements. At present,
there is no {\em a priori\/} method to determine the border indices of
neighboring blocks inside the chaotic layer. However, use of
classical information will warrant correct results to within an order
of magnitude. 

We note that our treatment is not limited to the case of an
$S$-matrix symmetry $S_{n,m}=S_{-n,-m}$ and could easily be extended
to non-symmetric systems (for example, an annular billiard with the
inner circle replaced by some non-symmetric shape),
or even to the case of tunneling at an accidental degeneracy between two
eigenphases. In the summation over paths, we merely require that the
initial and the final diagonal $S$-matrix elements are equal,
$S_{n_i,n_i}=S_{n_f,n_f}$. Note, however, that in the non-symmetric
case the contributions $\langle \bbox{n}_i|S^N|\bbox{n}_i\rangle$ and
$\langle \bbox{n}_f|S^N|\bbox{n}_f\rangle$ in
Eq.~(\ref{shift:iterates}) will in general not cancel. Their
difference is then likely to dominate the splitting. 

It is one of our main results to point out the importance of the beach
layer to the chaos-assisted tunneling phenomenon. The appearance of
classically chaotic, but not too unstable regions around regular
islands is generic in mixed systems. Such regions should always
support states if the mixing with the rest of phase space is
sufficiently slow (or if energy is sufficiently low). However, it must
be checked whether some of the importance of the beach layer should
actually be attributed to the tunneling ridge that favors tunneling
processes into the beach region (see Fig.~\ref{tunnel:fig}). Such a
test is given in our version of the Bohigas numerical experiment, see
Section \ref{dis:chaos:section}. 
We have verified that, at small $\delta$, there are {\em 
  no\/} visible tunneling ridges, and the tunneling amplitudes decay
monotonically away from the diagonal. Nevertheless
the beach region still governs the behavior of the eigenphase
splittings (see Fig.~\ref{dis:manyd39:fig}).  It must however be noted
that, in this case, the correspondence between
slow splitting modulations and the shift breaks down. The shift is
then determined by paths $n\ra n-1\ra n$ instead of paths leading to
the beach and back to $n$.   

It is another interesting point that the statistical results found in
Section \ref{Statistics} are independent of the explicit joint
distribution function of eigenphases, but can be derived under rather
general assumptions. We merely require that the joint distribution of
eigenphases vanishes at eigenphase degeneracies. 

This generality is extended even further by the observation that
enhancement of tunneling can also appear with help of {\em regular\/}
states. Indeed, in Section \ref{dis:chaos:section} we have even observed
the case of tunneling between a {\em chaotic\/} doublet via a resonant
regular state. This should serve as a reminder that only the phase
space {\em topology\/} determines the occurrence of tunneling, not its
regularity or chaoticity, and that chaos-assisted tunneling is, in
fact, a more general phenomenon of {\em transport\/}-assisted
tunneling. Additionally, the tunneling rate seems to be rather
insensitive to the rate of classical flux connecting the opposite
beach regions --- as long as there exists a classically allowed path
between them. When changing $\delta$ in Section
\ref{dis:chaos:section}, most of the tunneling enhancement was
related to the change of tunneling properties between the torus and
the beach region. Progressively rapid classical propagation across the
chaotic layer was related to tunneling enhancement of only one order
of magnitude --- out of five orders of magnitude in total
($\delta=0.07$--0.15). 

Our study of chaos-assisted tunneling has led to the most quantitative
treatment of the phenomenon to date. At the same time, some
challenging problems have been encountered. For example, 
we have seen that tunneling can occur between doublets  
localized on ``soft'' phase space structures such as the beach regions
or scarring periodic orbits. For these states, transport from one 
phase space structure to its symmetry-related partner is classically
allowed, but quantum mechanically forbidden. Apart from the intriguing
question about the quantum-mechanical localization mechanism giving
rise to these states, their doublet structure introduces additional
complications. For example, some of the doublets tunnel via resonant
processes, while others tunnel directly. A quantitative treatment
would certainly be desirable.

\subsection{Summary}

We studied dynamical tunneling between symmetry-related
phase space tori that are separated by a chaotic region. Using
scattering theory, we introduced a unitary matrix $S$ that constitutes
the quantum analogue of the classical Poincar\'e map. By expressing
eigenphase splittings and shifts in terms of matrix elements of high
iterates of $S$, we related these quantities to paths in phase
space. While paths contributing to the splitting connect the two
tunneling tori, paths that contribute to the shift lead from a
tunneling torus back to itself, leaving the torus at least once.  We
performed the summation over paths within a block-matrix
approximation, allocating different blocks to the two regular regions,
the chaotic sea, and the two intervening beach layers. Within this
approximation, we derived analytic expressions for the contributions
to the tunneling properties. Explicit inclusion of the beach blocks
enabled us to predict a number of new effects that could be verified for
the case of the annular billiard.  (I) As a function of an external
parameter the splitting varies on two scales: a rapid one attributed
to resonance denominators of regular and chaotic states, and a slow
one attributed to (squared) resonance denominators between regular
states and beach states. This diversity of scales is also observed in
statistical quantities, e.g~the distribution function
$P(\delta\theta)$ of eigenphase splittings $\delta\theta$. When
averaging over a sufficiently small range of a system parameter (such
that beach properties remain effectively constant), the splittings are
distributed with a Cauchy tail $P(\delta\theta)\sim
\delta\theta^{-2}$. When averaging over a large parameter range (such that 
beach resonances occur), the squared resonance denominators lead to a
$P(\delta\theta)\sim\delta\theta^{-3/2}$ power-law behavior.  
(II) Typically, the shift varies on the slow scale only and is much
larger than the splitting.

Analytical formulas at hand, we could also asses the relative
importance of tunneling amplitudes and classical transport properties
within the chaotic sea. As the annular billiard's eccentricity is
increased, most of the enhancement of tunneling rates can be
attributed to the tunneling amplitudes and resonances between regular
tori and the beach regions. Progressively faster classical transport
within the chaotic sea was found to play a minor role in the splitting
enhancement.

Finally, we derived the asymptotic form
$P(\delta\theta)\sim\delta\theta^{-2}$ of the splitting distribution's
large-$\delta\theta$ tail (average over a small parameter
range).  In this calculation, no explicit assumption about the form of
the joint distribution function of chaotic eigenphases was made;
it was merely required that the distribution is either Poissonian or
vanishes 
for degenerate eigenphases.  In order to give ``typical'' splitting
values, we calculated the median splitting by averaging over the
properties of the chaotic sea. Apart from an over-estimate
by an overall factor $\sim 5$, the 
predicted values for the median splittings closely follow the
numerical results over many orders of magnitude.

It is a pleasure to thank H.~A.~Weidenm\"uller for many conversations
and support. Also, we thank O.~Bohigas, F.~von Oppen,
S.~Tomsovic, and D.~Ullmo for discussions on several occasions. This
work was partially funded by a grant from MINERVA.   

\begin{appendix}
\section{Summation over Paths in the Splitting Formulas} 
\label{App:splittings}
In this Appendix we will derive formula (\ref{block:pathsum}) that
contains the contributions of the different families of paths to the
full sum over paths
\[
{\cal P}^N_{n,-n} = {\sum_{\{n\:\rightarrow
     -n\}}} \prod_{i=1}^{N-1}\:S_{\lambda_i,\lambda_{i+1}} 
\] 
leading from $n$ to $-n$. Recall that ${\cal P}^N_{n,-n}$ is related to the
splitting of the doublet $\delta\theta_n^{\pm}$ by
Eq.~(\ref{shift:splitting}).  As $N\sim\kappa/ \delta\theta_n$ is
taken to be large, we need only collect the leading-order contribution
in $N$. 
 
Let us consider the general case in which the $S$-matrix and
block-diagonalized into any number of blocks $S^{(i,j)}$,
$i,j=0,\ldots,K$. For the summation over paths, we merely require 
that the outermost couplings 
$S_{\lambda,\lambda'}^{(0,j)}$ and $S_{\lambda,\lambda'}^{(j,K)}$ be 
much smaller than entries of the internal coupling matrices. 
Suppose we want to collect the contributions from paths with $M+1$ steps
that start from $n$ and pass through the $M$ intermediate diagonal blocks
$S^{(i_1,i_1)},S^{(i_2,i_2)},\ldots,
S^{(i_{M},i_{M})}$ before arriving at $-n$, and let us for the moment
neglect repetitions in the block indices. 
It is equivalent to sum over all such paths in the
block-tridiagonal matrix 
\begin{eqnarray}
  \left(\begin{array}{cccccc}
    D_0   & C_0' &       &          &           &           \\
    C_0   & D_1   & C_1' &          &   0       &           \\
          & C_1   & D_2   & \ddots   &          &           \\
          &       &\ddots & \ddots   & C_{M-1}' &           \\
          &  0    &       & C_{M-1}  & D_{M}    & C_{M}'    \\
          &       &       &          & C_{M}    & D_0      
\end{array}
\right)\ ,
\label{app1:matrix}\end{eqnarray}
where each of the diagonal blocks 
$D_i=S_{\lambda,\lambda}^{(i,i)}\delta_{\lambda,\lambda'} $ is coupled
to its neighbor by the coupling block
$C_i=S_{\lambda,\lambda'}^{(i,i+1)}$. $C'$ is a short notation for 
the other coupling block, $C_i'=S_{\lambda',\lambda}^{(i+1,i)}$
(which need however not be the transpose of $C_i$, as $S$ is in
general not symmetric). 
$D_0=S_{n,n}=S_{-n,-n}$ contains the diagonal $S$-matrix element
associated with the tunneling tori.   

In addition to summations over the matrix indices of each block, 
we have to sum over
the staying times $N_0, \ldots, N_M$ inside the diagonal blocks
$D_0,\ldots,D_M$, respectively. Explicitly, we have to perform the sum
\begin{eqnarray}
  &&\sum_{\{N_0,\cdots N_M\}}\!\!
  D_0^{\,N_0}\,C_0\,D_1^{\,N_1}\cdots D_M^{N_M}\,C_{M}\,
   D_0^{\,N-N_0-N_1\ldots -N_M-M}
  \nonumber\\
  &&= \:D_0^{\,(N-M)}\!\!\sum_{\{N_0,\cdots N_M\}}\!\!
  C_0\,(D_1/D_0)^{N_1}\cdots (D_M/D_0)^{N_M} \,C_M\ ,
\nonumber\end{eqnarray} 
where the summations go over $N_M=0,\ldots,N-M$,
$N_{M-1}=0,\ldots,N-N_M-M$, and so on, up to
$N_0=N-N_M-N_{M-1}-\ldots-N_1-M$.
Summing over $N_0$ is trivial and generates a factor
$(N-N_M-\ldots-N_1-M+1)$. 
The remaining sums are then performed by repeated use of the formula
\begin{eqnarray}
  \sum_{\rho=0}^{R} (R-\rho+1)\,z^\rho=\frac{R+2}{1-z}
  + \frac{z^{R+2}-1}{(1-z)^2}\ ,
\label{app1:formula}\end{eqnarray}
For $z=D_i/D_0$ of absolute value $|z|<1$, we need only keep the
term $R/(1-z)$, because in further summations the remaining terms
generate sub-dominant contributions of ${\cal O}(1)$ or ${\cal
  O}(|z|^N)$. Neglecting these terms is
justified in our case, because we have assumed that the outermost
(tunneling) matrix elements are much smaller than internal transition
elements and by unitarity, $|D_i|<|D_0|$ for all $i\neq 0$.      
Keeping only the term $R/(1-z)$ of (\ref{app1:formula}),
the structure of the sum always remains the same, and each summation
results in a multiplicative factor \mbox{$(1-D_i/D_0)^{-1}$}. 
We arrive at the result that the sum over paths of length
$M$ passing through the blocks $D_0,\ldots,D_M$ is given as 
\begin{eqnarray}
{\cal P}_{n,-n}^{N\,(i_1,\ldots,i_M)} \sim N \:D_0^{\,(N-1)}\: C_0\:
\prod_{i=1}^{M}\:\left[\:\frac{1}{D_0-D_i}\,C_i\:\right]\ . 
\label{app1:pathsum}\end{eqnarray}

Eq.~(\ref{app1:pathsum}) was formulated without allowing for repetitions in
block indices and contains all coupling elements to lowest order. 
Loops in block index space that stay within
the inner blocks can however be included by allowing repeated indices
in Eq.~(\ref{app1:pathsum}). Paths with $k$ repetitions of the index
combination $(i,i+1)$, say, then give rise to contributions
containing a factor $[(D_0-D_i)^{-1} C_i' (D_0-D_{i+1})^{-1}C_i]^k$.   
The number of repetitions can be summed over, which leads to an
expression as in Eq.~(\ref{app1:pathsum}) with the replacement
\begin{eqnarray}
  &&\frac{1}{D_0-D_i}\:C_i
  \quad\mapsto\quad
  \nonumber\\
  &&\frac{1}{D_0-D_i}\:C_i\;\left[ 1- \frac{1}{D_0-D_{i+1}}\:C_i'\:
  \frac{1}{D_0-D_i}\:C_i\right]^{-1}\ .
\nonumber\end{eqnarray}
All types of loops can be included by the corresponding replacements,
giving rise to a continued fraction structure of ${\cal
  P}_{n,-n}^{N\,(i_1,\ldots,i_M)}$. 
\end{appendix}

\bibliographystyle{prsty}

\end{document}